%% file: paper.tex
\documentclass[12pt]{elsarticle}
\usepackage{lineno}

\usepackage{layouts}
\usepackage{graphics,graphicx}
\usepackage{pgfplots}
\usepackage{dcolumn}
\usepackage{bm}
\usepackage{subcaption}
\usepackage{setspace}

\usepackage{amsmath}
\usepackage{amssymb}
\newcommand\Rey{\mbox{\textit{Re}}}  
\usepackage{booktabs}

\begin{document}

\doublespacing
\begin{frontmatter}
\title{Can Minkowski tensors of a porous microstructure characterize its permeability? }

  \author[mss]{Prapanch Nair\corref{cor1}}
  \ead{prapanch@ipp.mpg.de}

  \author[mss]{Sebastian M\"uhlbauer}
  \author[iitd]{Shantanu Roy}
  \author[mss]{Thorsten P\"{o}schel}
  \address[mss]{Institute for Multiscale Simulation, Friedrich-Alexander University of Erlangen-Nuremberg, Germany.}
  \address[iitd]{Dept. of Chemical Engineering, Indian Institute of Technology Delhi, New Delhi, India.}

\date{\today}

\begin{abstract}
We show that the permeability of porous media can be reliably predicted from the Minkowski tensors (MTs) describing the pore microstructure geometry. 
To this end, we consider a large number of simulations of flow through periodic unit 
cells containing complex shaped obstacles. 
The prediction is achieved by training a deep neural network (DNN) using the simulation data with the MT elements as attributes. 
The obtained predictions allow for the conclusion that MTs of the pore microstructure contain sufficient information to determine the permeability, although the functional relation between the MTs and the permeability could be complex to determine. 

%
%
\end{abstract}


\end{frontmatter}
\section{Introduction}
The characterization of the microstructure of a porous medium is an unsolved
problem for nearly a century.
While the influence 
of the porosity on the permeability is well studied~\cite{Kozeny1927, Carman1937, costa2006permeability}, the effect of 
the shape of the porous microstructure itself on permeability has remained unclear. 
The difficulty is in identifying a shape descriptor that is concise yet relevant 
to the  permeability of the porous microstructure. 
Spatial discretizations (like voxel data from tomograms or the points on a reconstructed surface) or parametric representations (such as spherical harmonics) have been used as shape descriptors, traditionally. 
While the first class of descriptors depends on the spatial resolution and, therefore, may require processing a large amount of data, the latter 
does not apply to general porous structures. Simple representations of shape, relevant to 
flow through porous media, such as the widely used tortuosity~\cite{epstein1989tortuosity} or 
Euler characteristics~\cite{scholz2012permeability}, cannot fully characterize permeability. Fitting parameters, obtained experimentally, are required to relate permeability to 
these characterizations~\cite{ghanbarian2013tortuosity}.

We identify Minkowski tensors (MTs)~\cite{schroder2011minkowski} as shape descriptors for the porous microstructure by showing that MTs are sufficient to characterize its permeability. 
Minkowski tensors have been introduced~\cite{schroder2011minkowski} as a shape descriptor for a variety of different complex shaped media such as cellular structures and granular media. In the latter, for example, MTs reveal novel characteristics of the packings such as the isotropy and the angle of isotropy~\cite{schaller2016microscopic}. 
While Minkowski functionals (MTs are a generalization of the functionals ~\cite{schrturk10-1}) were proposed for the characterization of porous permeability~\cite{armstrong2019porous}, until now there is no evidence that MTs can reliably characterize the permeability~\cite{scholz2012permeability, slotte2020predicting}. 

In this study, we simulate the flow around an obstacle of a randomly generated, two-dimensional, simply-connected shape placed within a periodic cell. Data from 
a plethora
of such flow simulations each corresponding to a random shape is used to train a deep neural network (DNN) to  show that the Minkowski tensors can predict permeability to very high accuracy. Thus, we demonstrate that the MTs for a given structure are sufficient to characterize  the 
permeability for the 2D periodic porous media considered here. 
Previous studies (for example~\cite{slotte2020predicting}) have shown that there is no linear relationship 
between Minkowski functionas and the porous permeability.
Since DNNs are known to capture complex non-linear relations, they can be employed to show if
permeability is related to the MTs of the microstructure. 

In the current paper, we show that Minkowski tensors are pertinent descriptors of 
the permeability through a data science experiment. We generate pseudo-random shapes, simulate fluid flow 
around these shapes, and compute the permeability for each of the shapes. Finally 
we train a DNN to show that the MTs, used as input features, can predict 
the computed permeabilities. 

\section{Shape generation and flow simulation}

We simulate the flow through representative elemental volumes (REVs) of porous media.
The solid walls within the REVs need to have shapes that are complex enough such that 
their influence on the permeability cannot be predicted by the porosity alone ~\cite{Kozeny1927,yazdchi2011microstructural}. Also, the flow domain needs to be simple enough such that it is feasible to generate
a large amount of data for the DNN's training.
Based on these considerations, we generate square 2D domains with periodic boundary
conditions containing a simply connected arbitrarily 
shaped obstacle of invariant area. The obstacle is assigned no-slip wall boundary condition for the flow simulation.
Each such domain thus forms the representative elemental volume (REV) for a separate porous medium. Such 
systems belong to the category of fibrous porous media and are widely considered in literature~\cite{nabovati2009general} for fundamental studies on permeability.

\begin{figure}[htb]
  \centering
  \input{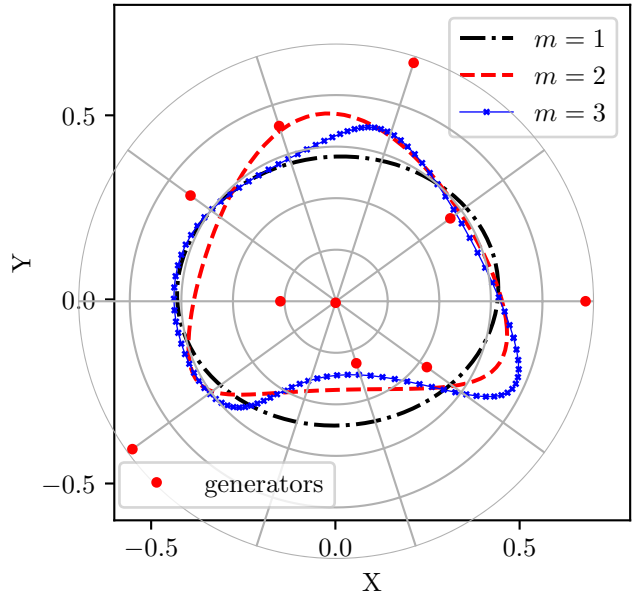}
  \caption{Examples of shapes generated by parametric curves according to Eq. (\ref{eq:shape}) to points located
  at random radial locations}
  \label{fig:shape}
\end{figure}

Figure \ref{fig:shape} illustrates the generation of obstacle shapes. Ten points 
are spawned at equal angular intervals with randomly generated radial coordinates. 
A Fourier parametric curve is fitted to these points using least square minimization. Similar approaches are widely used in reconstructing arbitrary non-convex shapes~\cite{wei2018simple,yotter2011topological}. 
The parametric representation of the shape of the obstacle is of the form:

\begin{align}
  x(\theta) &= \frac{1}{2}X_0 \Sigma_{i=0}^m X_i^c \cos ({i}\theta) + X_i^s \sin (i\theta) \\ 
  y(\theta) &= \frac{1}{2}Y_0 \Sigma_{i=0}^m Y_i^c \cos ({i}\theta) + Y_i^s \sin (i\theta)  .
  \label{eq:shape}
\end{align}

For $\theta \in \left[0,2\pi \right)$, $x$ and $y$ describe a shape of order $m$. 
The coefficients $X_0/2$ and $Y_0/2$ are the coordinates of the centroid of a shape. 
The area of a given shape can be computed by:
\begin{equation}
  A = \int\limits_0^{2\pi} y(\theta) x'(\theta)d\theta,
  \label{eq:area}
\end{equation}
where  $x' = dx/d\theta$. The shapes 
are scaled such that the area is maintained to an invariant value of 0.5 
for all shapes (to keep the porosity fixed: $\phi = 0.5$) and the centroid $(X_0/2,Y_0/2)$
is translated to the center of the unit square domain. Shapes with regions 
outside the square, occurrence of sharp spikes and 
self intersections are eliminated during the data generation. Different
values for $m$, namely $1$, $2$ and $3$, are considered, and for each of these 
orders ($m$), 100,000 shapes are generated. 
For $m=1$, only ellipses are obtained, while higher values of $m$ result in shapes with more undulations. 
Figure \ref{fig:shape}(b) shows shapes of different order generated with the same 
set of generator points. 

The flow through each of the REVs is then simulated using the finite volume method
\cite{popinet2003gerris}. The solver \cite{popinet2003gerris}
uses an adaptive quadtree mesh 
which eliminates the need for 
additional meshing and pre-processing difficulties. For each REV, three different 
pressure gradient values are applied to drive the flow in the horizontal 
direction. 
The largest pressure gradient magnitude
is chosen such that the corresponding Reynolds number is about 1 ($\Rey =\rho U r/\mu$, where $r$ is the radius of the 
circle of equal area as the obstacle, $U$ is the superficial velocity and $\rho$ and $\mu$ are the density and viscosity of the fluid). 
Therefore, we expect the flow regime to not deviate substantially from the Darcy regime at least
at the smallest pressure gradient magnitude considered.
Darcy's equation for porosity reads:
\begin{equation}
  \frac{dp}{dx}=-\frac{\mu}{K}U.
  \label{eq:darcy}
\end{equation}
Here $dp/dx$ is the applied pressure gradient in the $x$-direction, $U$ is the 
obtained superficial 
velocity in $x$-direction, $\mu$ is the viscosity, and $K$ is the permeability (assuming an isotropic 
porous medium). For each REV, we fit the pressure gradient,  $dp / dx$, to a $2^\text{nd}$ order polynomial 
function of the superficial velocity, $U$:

\begin{equation}
  \frac{dp}{dx}=-\frac{\mu}{K}U-\frac{\rho}{K_1}U^2 ,
  \label{eq:forchheimer}
\end{equation}
where $\rho$ is the density of the fluid and $K_1$ is the inertial permeability, also known as the Forchheimer permeability \cite{whitaker1996forchheimer}. The slope of this curve at the 
origin ($U=0$), gives the coefficient of the linear term in $U$, namely 
$\mu/K$. 
For a given viscosity, $\mu$, the value of $K$ can be thus computed for each REV from the simulated flowfield. 
\begin{figure}[htb]
  \centering
\begin{subfigure}[b]{0.49\textwidth}
  \includegraphics[width=\textwidth]{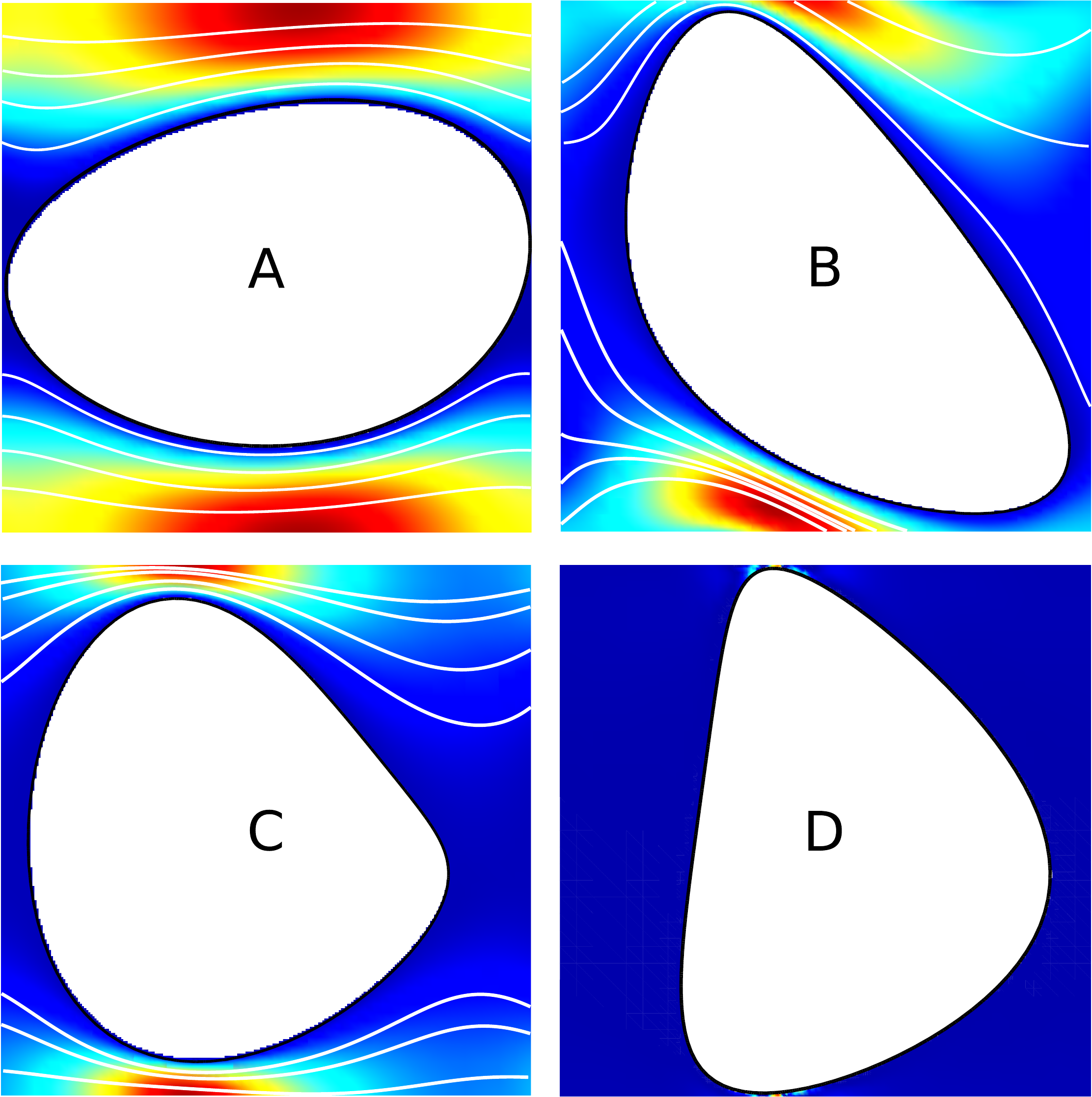}
  \caption{}
\end{subfigure}
  \begin{subfigure}[b]{0.49\textwidth}
  \input{four_shapes_fit.pgf}
  \caption{}
\end{subfigure}
\caption{Flow through four exemplary shapes for $m=2$: (A) and (D) correspond to the shapes with maximum and minimum permeability in the dataset, respectively. The contours in (a) correspond to the velocity component in $x$-direction and the white lines represent the flow streamlines. The plot (b) shows the fit of Eq. \ref{eq:forchheimer} for the velocity values (normalised as Re) for varying pressure gradient (in $x$-direction). The solid line denotes the circular shape used as a reference shape. }
  \label{fig:four_shapes}
\end{figure}

Figure \ref{fig:four_shapes}a shows four example shapes from the dataset including the 
extreme cases, i.e., the most permeable REV (A) as well as the least permeable REV (D)
among the shapes with $m=2$.
The plot in Fig. \ref{fig:four_shapes}b shows the Reynolds number  
against the pressure gradient for each of these example shapes and the polynomial fit to these data.
\begin{figure}[htb]
  \centering
  \input{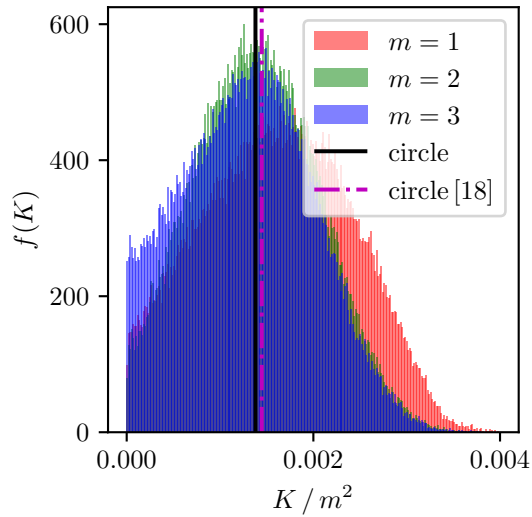}
  \caption{The probability density, $f$, of permeability, $ K$, for the shapes of different complexity, $m$, with porosity  $\phi=0.5$. For each order parameter, $m$, $10^5$ shapes were used. The solid line indicates the permeability for 
  a circular obstacle from simulation and the dashed line shows the permeability for a circle determined analytically~\cite{Bruschke1993}.}
 
  \label{fig:permdist}
\end{figure}

Figure \ref{fig:permdist} shows the distribution of the permeability values for each set of shapes with the corresponding values of $m$. As $m$ increases, we observe that an increasing number of shapes in the dataset has permeability zero. 
This is because the probability of a shape blocking the flow 
increases as the number of free parameters determining the shape increases. 
The permeability values with a circular obstacle ($\phi=0.5$) in the periodic domain,  
calculated analytically~\cite{Bruschke1993} and obtained from simulations are marked using a dash-dot line and a solid line, respectively. The permeability of the circular obstacle is close to the value corresponding to the peak probability density  of our dataset.  

\section{Minkowski tensors}
Minkowski tensors are generalizations of the Minkowski functionals~\cite{schroder2011minkowski,armstrong2019porous}
and can characterize a shape using translation covariant and translation invariant tensors. Four linearly independent 
Minkowski tensors can be defined for two spatial dimensions and six for three spatial dimensions.
Our study includes only two spatial dimensions. The MTs are integrals over shapes of area $\Omega$ and boundary $\partial \Omega$:
\begin{align}
  W_0^{2,0} &= \int\limits_\Omega \mathbf{r} \otimes \mathbf{r}\, dA \label{eq:mt1}\\
  W_1^{2,0} &= \frac{1}{2}\int\limits_{\partial \Omega} \mathbf{r} \otimes \mathbf{r}\, dr \label{eq:mt2} \\
  W_2^{2,0} &= \frac{1}{2}\int\limits_{\partial \Omega} \kappa (\mathbf{r}) \mathbf{r} \otimes \mathbf{r}\, dr \label{eq:mt3} \\
  W_1^{0,2} &= \frac{1}{2}\int\limits_{\partial \Omega}  \mathbf{n} \otimes \mathbf{n}\, dr \label{eq:mt4}.
\end{align}
Here, the vectors $\mathbf{r}$ and $\mathbf{n}$ are the position
and the unit normal vector at the surface, respectively, and $\kappa$ is the local curvature. 
We use a software that employs an efficient numerical method to compute the MTs~\cite{schrturk10-1}.

Each MT is a symmetric, two-dimensional, second order tensor and, thus, 
contains 3 independent scalar values. For the purpose of building the DNN model 
we use the Eigenvalues as well as the 
elements of the Eigenvectors as training features. 
This results in six features (scalar values) for each MT, making a total of 24 features.

\section{The DNN model}


The features (Eigenvalues and Eigenvector elements) are provided as input nodes to a DNN comprised of 7 hidden layers and an output layer with a single node representing the permeability value. We have chosen the \emph{tanh} function as the activation function. 
We have used the \emph{Keras}~\cite{Charles2013} library with a \emph{TensorFlow} backend for the purpose of setting up the DNN and training it.
Further details of the neural network 
architecture used in our study is provided in Table \ref{tab:parameters}.
The $L_2$ norm of the difference between the predicted  and  the actual $K$ value is used as the \emph{loss function} for the training.
The DNN is trained by optimizing the weights and biases of each node (neuron) of the network for minimizing the loss function through several iterations (also known as epochs).  

\begin{table}
  \centering
  \begin{tabular}{ll}
    \toprule
    Parameter & Value \\
    \midrule
    activation function & $\tanh$ \\
    hidden layer type & dense \\
    number of layers & 9 \\
    hidden layer sizes & $\{24,\,24,\,48,\,72,\,32,\,4,\,1\}$ \\
    loss function & mean squared error \\
    optimizer & stochastic gradient descent \\
    initial learning rate & 0.005 \\
    feature normalisation & $\min-\max$ \\ 
    \bottomrule
  \end{tabular}
  \caption{Training parameters used in training the DNN. For details of implementation, refer to the scripts in the supplementary material.}
  \label{tab:parameters}
\end{table}


The available dataset is split into a \emph{training set} containing 95\% of the data selected at random and a \emph{validation set} containing the rest of the data.
While the training set is used to optimize the nodal weights, the validation set 
is used to establish the accuracy of the prediction and to check for overfitting. 
A higher rate of decrease in error for the training 
set than that for the validation set would mean overfitting of the training data. We stop the training process either when the error stops decreasing simultaneously for both the training and validation sets or when overfitting sets in. The training parameters were chosen by experimenting with the training process.   


We show, in Fig.\ref{fig:learning}, that the model we trained predicts K 
within 6\,\%  mean squared error (both training and validation dataset) relative to the total variance 
in the permeability values for each of the set of shapes. 
It may be possible to improve the parameters of the DNN and achieve 
better prediction of K. However, the achieved accuracy of prediction is sufficient to show that MTs effectively characterize the permeability.

\section{Results and discussion}

Beyond the obvious dependence of permeability on porosity \cite{Kozeny1927}, a comprehensive dependence of 
permeability on shape has been a subject of intense research~\cite{bentz2000microstructure,tamayol2011transverse,sun2011multiscale,noiriel2004investigation,garboczi1990permeability}. 
Previous studies that used simple fit functions were unable to conclude that 
the permeability is dependent on MTs~\cite{armstrong2019porous,slotte2020predicting}.
A DNN, on the other hand, is able to learn non-linear relationships, and therefore, can reveal if
Minkowski tensors characterize the permeability.  

\begin{figure}[htb]
  \centering
  \begin{subfigure}[b]{0.32\textwidth}
    \includegraphics[width=\textwidth]{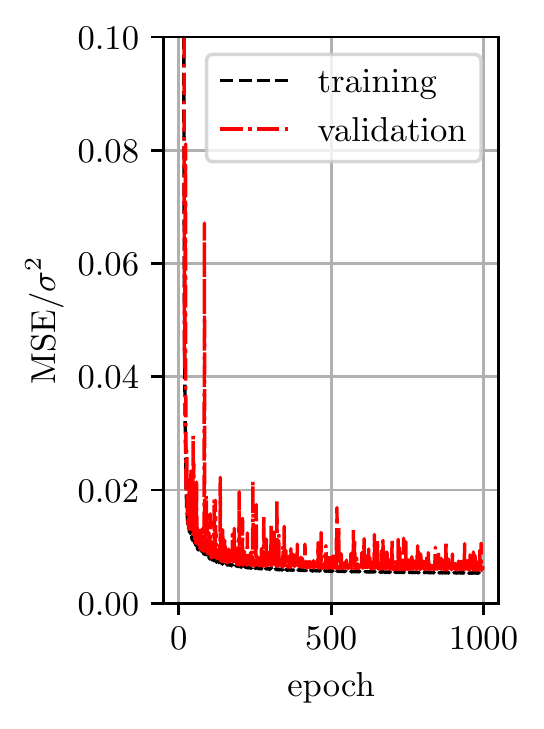}
    \caption{$m=1$}
  \end{subfigure}
  \begin{subfigure}[b]{0.32\textwidth}
    \includegraphics[width=\textwidth]{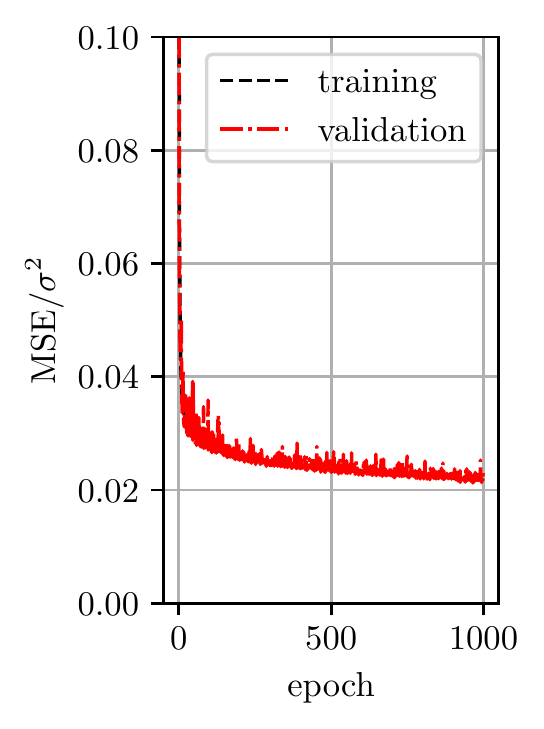}
    \caption{$m=2$}
    \label{fig:learn_m2}
  \end{subfigure}
  \begin{subfigure}[b]{0.32\textwidth}
    \includegraphics[width=\textwidth]{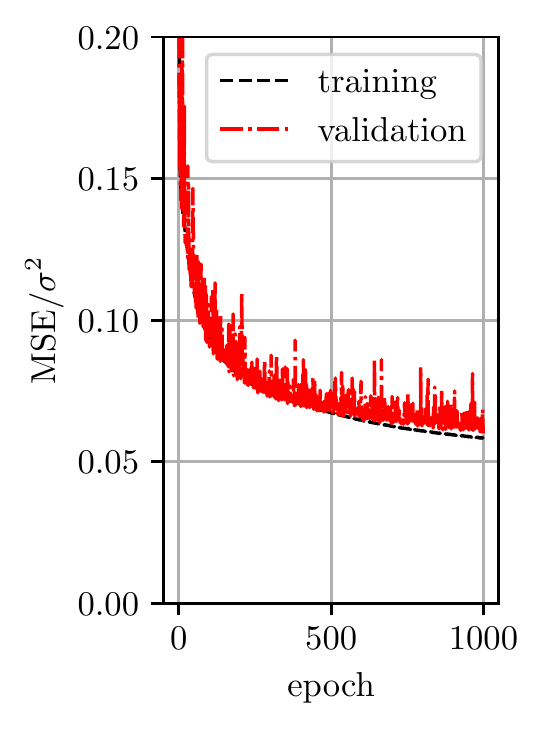}
    \caption{$m=3$}
  \end{subfigure}
  \caption{Training and validation of the DNN model for shapes of different complexity ($m$). The Y-axis in each of the plots shows the mean-squared error (MSE) of the model prediction normalized by the variance in the permeability values in each case. As test set we randomly take about 10\,\% out of the total dataset in each case. The X-axis shows the number of training epochs (iterations).}
  \label{fig:learning}
\end{figure}

Figure \ref{fig:learning} shows the convergence of the mean squared error (MSE) normalized by the total variance of the permeability values of all the shapes, with increasing epochs. While the training converges for $m=1$, the DNN starts overfitting for the shapes with $m=2$ and $m=3$ and hence the training was terminated when the error in the training set started deviating from the error in the validation set. The training parameters used were kept constant across these data sets and the same parameters do not result in same accuracy across the data sets. We show that training accuracy is the highest ($\textrm{MSE}/\sigma^2 < 1\%$) for shapes with $m=1$ (ellipses) and the error increases with increasing $m$. Nevertheless, the overall accuracy is still high enough for 
practical purposes. 


\begin{figure}[htb]
  \centering
  \begin{subfigure}[b]{0.49\textwidth}
    \includegraphics[width=\textwidth]{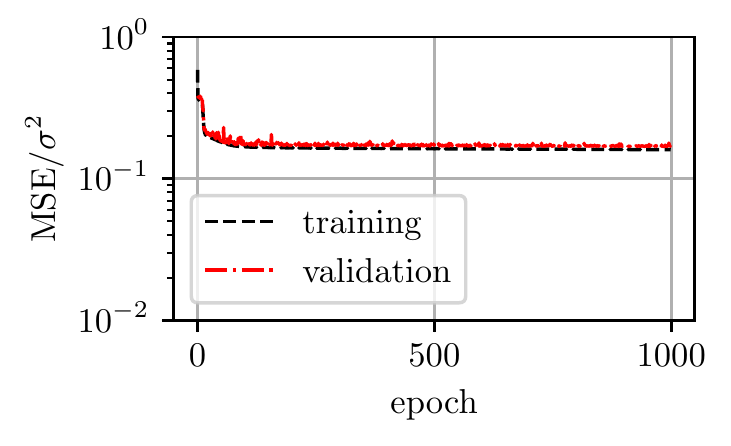}
    \caption{$W_0^{2,0}$}
  \end{subfigure}
  \begin{subfigure}[b]{0.49\textwidth}
    \includegraphics[width=\textwidth]{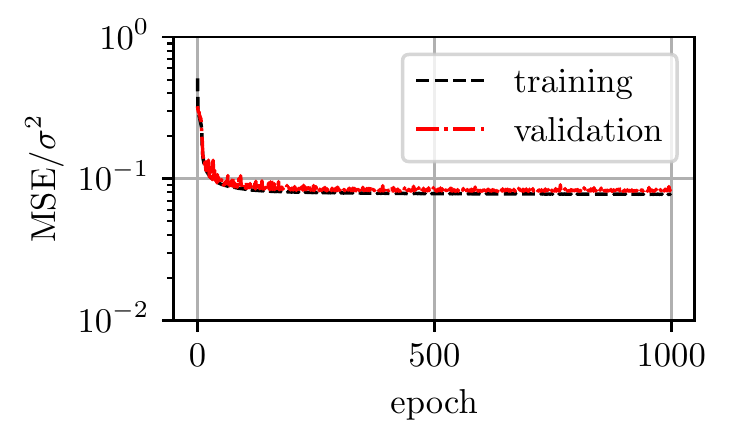}
    \caption{$W_1^{2,0}$}
  \end{subfigure}
  \begin{subfigure}[b]{0.49\textwidth}
    \includegraphics[width=\textwidth]{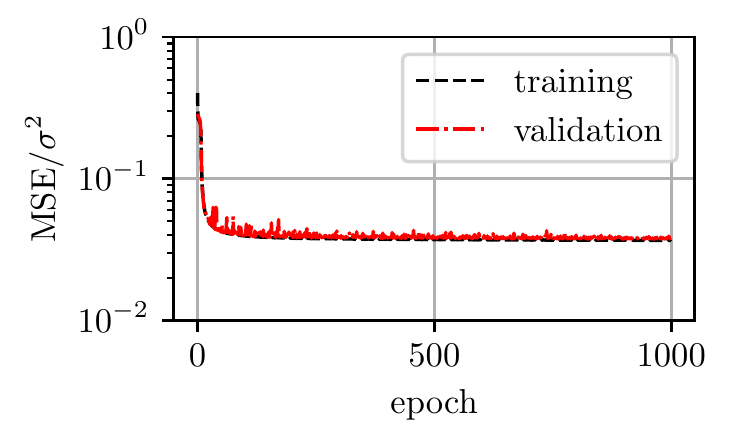}
    \caption{$W_2^{2,0}$}
  \end{subfigure}
  \begin{subfigure}[b]{0.49\textwidth}
    \includegraphics[width=\textwidth]{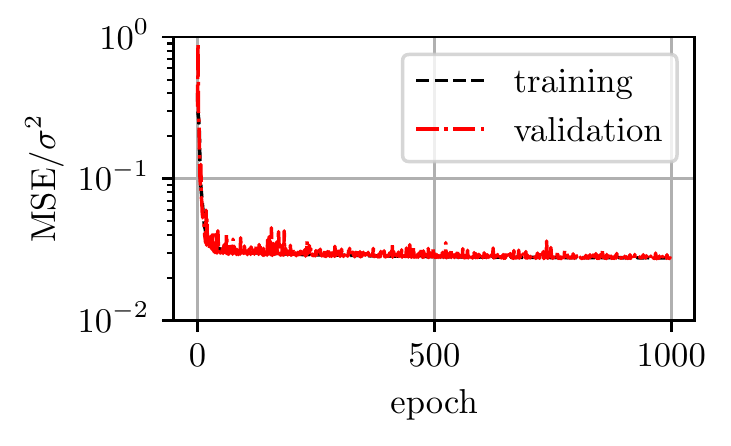}
    \caption{$W_1^{0,2}$}
  \end{subfigure} 
  \caption{Training and validation of the DNN with different Minkowski tensors used exclusively as features. }
  \label{fig:learn_mt}
\end{figure}

In other applications using MTs such as in the structure of sphere packings, the 
higher order MTs have not revealed additional insights about the structures than 
those by the lower order MTs~\cite{schaller2016microscopic}, raising the question 
whether these higher order measures are significant for characterization of the permeability. 
The volume moment tensor $W_0^{2,0}$ is equivalent to the area moment of inertia tensor and could thus 
capture effects such as blocking of the flow due to orientation of shapes as is evident from several other studies~\cite{sobera2006hydraulic,chen2008transverse}. 
In order to show that all the MTs contribute to the 
prediction of the permeability,
 we present training results where only one MT is used 
 at a time (6 features) considering the dataset for $m=2$, in Fig. \ref{fig:learn_mt}. For the Y-axis in Fig. \ref{fig:learn_mt},  the same data 
and normalization approach were used as in Fig. \ref{fig:learn_m2}. 
Clearly, all four MTs are independently able to predict the permeability to great
accuracy ($\textrm{MSE}/\sigma^2<10\%$) and, thus, contribute to the model. Thus MTs embody much more information about the shape, relevant to the permeability, than simple descriptions such as \emph{tortuosity}~\cite{ghanbarian2013tortuosity} and porosity \cite{yazdchi2011microstructural}. 

\begin{figure}[htb]
  \centering
  \input{convergence.pgf}
  \caption{The minimum error in validation data obtained for different training data sizes for the shapes of order $m=2$. Each point corresponds to a different DNN model obtained with a data size shown on the $x$-axis. The same training parameters were used. The $y$-axis shows the mean square error of permeability estimate normalized with total variance in the data for permeability values. }
  \label{fig:learning_size}
\end{figure}
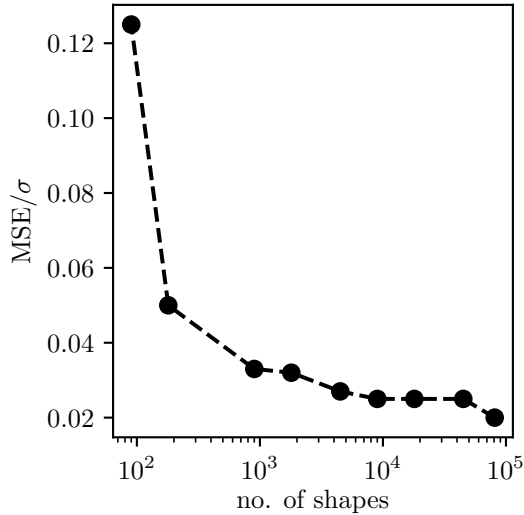

It is a worthwhile question to ask whether a large amount of 
data may always be required for a successful permeability prediction model based on the MTs. 
To answer this question we perform the training 
process with different training data set sizes and juxtapose these results in 
Fig. \ref{fig:learning_size}.
Here we plot the minimum error achieved against the size of the training data ($m=2$).
Remarkably, only a very small number of samples, namely 100, is required to predict 
the permeability ($\textrm{MSE}/\sigma^2<10\%$). This 
suggests that a similar exercise using three-dimensional domains can be achieved with 
computationally feasible data size. 

\section{Conclusion}
We have demonstrated that the answer to the titular question whether Minkowski tensors
of a porous microstructure can be related exclusively to the permeability of the 
medium is a `yes.' We have generated a large enough dataset of complex shapes 
of constant porosity with 
different orders of complexity and performed CFD simulations of flows through 
the REVs containing these non-convex simply connected shapes. 
We trained a DNN using these simulated data, validated the model, and 
demonstrated its predictive accuracy for different shape complexity and for different dataset
sizes. Reasonably accurate prediction results from even a small number ($\approx 100$) of data points for
training the model. This indicates that  
computationally more challenging systems can be investigated following the procedure 
outlined above.

Physically, these unit cells with obstacles
represent cross sections of uni-directional regular fibrous porous media 
and may seem limited in application. However, the 
motivation to the above exercise has been to identify a simple system where a 
data experiment relating the MTs to the permeability may be conducted. Our results 
show that Minkowski tensors are adequate descriptors of the pore microstructure. This observation motivates the hypothesis that
the same may be true for much larger scale systems in three dimensions 
and involving non-uniform porous microstructure. Further investigations into 
anisotropic porous media, where the tensorial 
nature of permeability is important is a natural extension to the present work. 





\input{paper.bbl}
\end{document}

%% file: four_shapes_fit.pgf
\begingroup%
\makeatletter%
\begin{pgfpicture}%
\pgfpathrectangle{\pgfpointorigin}{\pgfqpoint{2.665278in}{3.010926in}}%
\pgfusepath{use as bounding box, clip}%
\begin{pgfscope}%
\pgfsetbuttcap%
\pgfsetmiterjoin%
\definecolor{currentfill}{rgb}{1.000000,1.000000,1.000000}%
\pgfsetfillcolor{currentfill}%
\pgfsetlinewidth{0.000000pt}%
\definecolor{currentstroke}{rgb}{1.000000,1.000000,1.000000}%
\pgfsetstrokecolor{currentstroke}%
\pgfsetdash{}{0pt}%
\pgfpathmoveto{\pgfqpoint{0.000000in}{0.000000in}}%
\pgfpathlineto{\pgfqpoint{2.665278in}{0.000000in}}%
\pgfpathlineto{\pgfqpoint{2.665278in}{3.010926in}}%
\pgfpathlineto{\pgfqpoint{0.000000in}{3.010926in}}%
\pgfpathclose%
\pgfusepath{fill}%
\end{pgfscope}%
\begin{pgfscope}%
\pgfsetbuttcap%
\pgfsetmiterjoin%
\definecolor{currentfill}{rgb}{1.000000,1.000000,1.000000}%
\pgfsetfillcolor{currentfill}%
\pgfsetlinewidth{0.000000pt}%
\definecolor{currentstroke}{rgb}{0.000000,0.000000,0.000000}%
\pgfsetstrokecolor{currentstroke}%
\pgfsetstrokeopacity{0.000000}%
\pgfsetdash{}{0pt}%
\pgfpathmoveto{\pgfqpoint{0.461111in}{0.499691in}}%
\pgfpathlineto{\pgfqpoint{2.553611in}{0.499691in}}%
\pgfpathlineto{\pgfqpoint{2.553611in}{2.764691in}}%
\pgfpathlineto{\pgfqpoint{0.461111in}{2.764691in}}%
\pgfpathclose%
\pgfusepath{fill}%
\end{pgfscope}%
\begin{pgfscope}%
\pgfsetbuttcap%
\pgfsetroundjoin%
\definecolor{currentfill}{rgb}{0.000000,0.000000,0.000000}%
\pgfsetfillcolor{currentfill}%
\pgfsetlinewidth{0.803000pt}%
\definecolor{currentstroke}{rgb}{0.000000,0.000000,0.000000}%
\pgfsetstrokecolor{currentstroke}%
\pgfsetdash{}{0pt}%
\pgfsys@defobject{currentmarker}{\pgfqpoint{0.000000in}{-0.048611in}}{\pgfqpoint{0.000000in}{0.000000in}}{%
\pgfpathmoveto{\pgfqpoint{0.000000in}{0.000000in}}%
\pgfpathlineto{\pgfqpoint{0.000000in}{-0.048611in}}%
\pgfusepath{stroke,fill}%
}%
\begin{pgfscope}%
\pgfsys@transformshift{0.556225in}{0.499691in}%
\pgfsys@useobject{currentmarker}{}%
\end{pgfscope}%
\end{pgfscope}%
\begin{pgfscope}%
\pgftext[x=0.556225in,y=0.402469in,,top]{\rmfamily\fontsize{10.000000}{12.000000}\selectfont \(\displaystyle 0\)}%
\end{pgfscope}%
\begin{pgfscope}%
\pgfsetbuttcap%
\pgfsetroundjoin%
\definecolor{currentfill}{rgb}{0.000000,0.000000,0.000000}%
\pgfsetfillcolor{currentfill}%
\pgfsetlinewidth{0.803000pt}%
\definecolor{currentstroke}{rgb}{0.000000,0.000000,0.000000}%
\pgfsetstrokecolor{currentstroke}%
\pgfsetdash{}{0pt}%
\pgfsys@defobject{currentmarker}{\pgfqpoint{0.000000in}{-0.048611in}}{\pgfqpoint{0.000000in}{0.000000in}}{%
\pgfpathmoveto{\pgfqpoint{0.000000in}{0.000000in}}%
\pgfpathlineto{\pgfqpoint{0.000000in}{-0.048611in}}%
\pgfusepath{stroke,fill}%
}%
\begin{pgfscope}%
\pgfsys@transformshift{1.163246in}{0.499691in}%
\pgfsys@useobject{currentmarker}{}%
\end{pgfscope}%
\end{pgfscope}%
\begin{pgfscope}%
\pgftext[x=1.163246in,y=0.402469in,,top]{\rmfamily\fontsize{10.000000}{12.000000}\selectfont \(\displaystyle 1\)}%
\end{pgfscope}%
\begin{pgfscope}%
\pgfsetbuttcap%
\pgfsetroundjoin%
\definecolor{currentfill}{rgb}{0.000000,0.000000,0.000000}%
\pgfsetfillcolor{currentfill}%
\pgfsetlinewidth{0.803000pt}%
\definecolor{currentstroke}{rgb}{0.000000,0.000000,0.000000}%
\pgfsetstrokecolor{currentstroke}%
\pgfsetdash{}{0pt}%
\pgfsys@defobject{currentmarker}{\pgfqpoint{0.000000in}{-0.048611in}}{\pgfqpoint{0.000000in}{0.000000in}}{%
\pgfpathmoveto{\pgfqpoint{0.000000in}{0.000000in}}%
\pgfpathlineto{\pgfqpoint{0.000000in}{-0.048611in}}%
\pgfusepath{stroke,fill}%
}%
\begin{pgfscope}%
\pgfsys@transformshift{1.770266in}{0.499691in}%
\pgfsys@useobject{currentmarker}{}%
\end{pgfscope}%
\end{pgfscope}%
\begin{pgfscope}%
\pgftext[x=1.770266in,y=0.402469in,,top]{\rmfamily\fontsize{10.000000}{12.000000}\selectfont \(\displaystyle 2\)}%
\end{pgfscope}%
\begin{pgfscope}%
\pgfsetbuttcap%
\pgfsetroundjoin%
\definecolor{currentfill}{rgb}{0.000000,0.000000,0.000000}%
\pgfsetfillcolor{currentfill}%
\pgfsetlinewidth{0.803000pt}%
\definecolor{currentstroke}{rgb}{0.000000,0.000000,0.000000}%
\pgfsetstrokecolor{currentstroke}%
\pgfsetdash{}{0pt}%
\pgfsys@defobject{currentmarker}{\pgfqpoint{0.000000in}{-0.048611in}}{\pgfqpoint{0.000000in}{0.000000in}}{%
\pgfpathmoveto{\pgfqpoint{0.000000in}{0.000000in}}%
\pgfpathlineto{\pgfqpoint{0.000000in}{-0.048611in}}%
\pgfusepath{stroke,fill}%
}%
\begin{pgfscope}%
\pgfsys@transformshift{2.377287in}{0.499691in}%
\pgfsys@useobject{currentmarker}{}%
\end{pgfscope}%
\end{pgfscope}%
\begin{pgfscope}%
\pgftext[x=2.377287in,y=0.402469in,,top]{\rmfamily\fontsize{10.000000}{12.000000}\selectfont \(\displaystyle 3\)}%
\end{pgfscope}%
\begin{pgfscope}%
\pgftext[x=1.507361in,y=0.223457in,,top]{\rmfamily\fontsize{10.000000}{12.000000}\selectfont Re}%
\end{pgfscope}%
\begin{pgfscope}%
\pgfsetbuttcap%
\pgfsetroundjoin%
\definecolor{currentfill}{rgb}{0.000000,0.000000,0.000000}%
\pgfsetfillcolor{currentfill}%
\pgfsetlinewidth{0.803000pt}%
\definecolor{currentstroke}{rgb}{0.000000,0.000000,0.000000}%
\pgfsetstrokecolor{currentstroke}%
\pgfsetdash{}{0pt}%
\pgfsys@defobject{currentmarker}{\pgfqpoint{-0.048611in}{0.000000in}}{\pgfqpoint{0.000000in}{0.000000in}}{%
\pgfpathmoveto{\pgfqpoint{0.000000in}{0.000000in}}%
\pgfpathlineto{\pgfqpoint{-0.048611in}{0.000000in}}%
\pgfusepath{stroke,fill}%
}%
\begin{pgfscope}%
\pgfsys@transformshift{0.461111in}{0.602646in}%
\pgfsys@useobject{currentmarker}{}%
\end{pgfscope}%
\end{pgfscope}%
\begin{pgfscope}%
\pgftext[x=0.294444in,y=0.554420in,left,base]{\rmfamily\fontsize{10.000000}{12.000000}\selectfont \(\displaystyle 0\)}%
\end{pgfscope}%
\begin{pgfscope}%
\pgfsetbuttcap%
\pgfsetroundjoin%
\definecolor{currentfill}{rgb}{0.000000,0.000000,0.000000}%
\pgfsetfillcolor{currentfill}%
\pgfsetlinewidth{0.803000pt}%
\definecolor{currentstroke}{rgb}{0.000000,0.000000,0.000000}%
\pgfsetstrokecolor{currentstroke}%
\pgfsetdash{}{0pt}%
\pgfsys@defobject{currentmarker}{\pgfqpoint{-0.048611in}{0.000000in}}{\pgfqpoint{0.000000in}{0.000000in}}{%
\pgfpathmoveto{\pgfqpoint{0.000000in}{0.000000in}}%
\pgfpathlineto{\pgfqpoint{-0.048611in}{0.000000in}}%
\pgfusepath{stroke,fill}%
}%
\begin{pgfscope}%
\pgfsys@transformshift{0.461111in}{1.631091in}%
\pgfsys@useobject{currentmarker}{}%
\end{pgfscope}%
\end{pgfscope}%
\begin{pgfscope}%
\pgftext[x=0.294444in,y=1.582866in,left,base]{\rmfamily\fontsize{10.000000}{12.000000}\selectfont \(\displaystyle 1\)}%
\end{pgfscope}%
\begin{pgfscope}%
\pgfsetbuttcap%
\pgfsetroundjoin%
\definecolor{currentfill}{rgb}{0.000000,0.000000,0.000000}%
\pgfsetfillcolor{currentfill}%
\pgfsetlinewidth{0.803000pt}%
\definecolor{currentstroke}{rgb}{0.000000,0.000000,0.000000}%
\pgfsetstrokecolor{currentstroke}%
\pgfsetdash{}{0pt}%
\pgfsys@defobject{currentmarker}{\pgfqpoint{-0.048611in}{0.000000in}}{\pgfqpoint{0.000000in}{0.000000in}}{%
\pgfpathmoveto{\pgfqpoint{0.000000in}{0.000000in}}%
\pgfpathlineto{\pgfqpoint{-0.048611in}{0.000000in}}%
\pgfusepath{stroke,fill}%
}%
\begin{pgfscope}%
\pgfsys@transformshift{0.461111in}{2.659537in}%
\pgfsys@useobject{currentmarker}{}%
\end{pgfscope}%
\end{pgfscope}%
\begin{pgfscope}%
\pgftext[x=0.294444in,y=2.611312in,left,base]{\rmfamily\fontsize{10.000000}{12.000000}\selectfont \(\displaystyle 2\)}%
\end{pgfscope}%
\begin{pgfscope}%
\pgftext[x=0.238889in,y=1.632191in,,bottom,rotate=90.000000]{\rmfamily\fontsize{10.000000}{12.000000}\selectfont \(\displaystyle dp/dx\, /\, Nm^{-3} \)}%
\end{pgfscope}%
\begin{pgfscope}%
\pgftext[x=0.461111in,y=2.806358in,left,base]{\rmfamily\fontsize{10.000000}{12.000000}\selectfont \(\displaystyle \times10^{-9}\)}%
\end{pgfscope}%
\begin{pgfscope}%
\pgfpathrectangle{\pgfqpoint{0.461111in}{0.499691in}}{\pgfqpoint{2.092500in}{2.265000in}} %
\pgfusepath{clip}%
\pgfsetrectcap%
\pgfsetroundjoin%
\pgfsetlinewidth{1.505625pt}%
\definecolor{currentstroke}{rgb}{0.000000,0.000000,0.000000}%
\pgfsetstrokecolor{currentstroke}%
\pgfsetdash{}{0pt}%
\pgfpathmoveto{\pgfqpoint{0.556225in}{0.602646in}}%
\pgfpathlineto{\pgfqpoint{0.890309in}{1.116868in}}%
\pgfpathlineto{\pgfqpoint{1.137086in}{1.631091in}}%
\pgfpathlineto{\pgfqpoint{1.538610in}{2.659537in}}%
\pgfusepath{stroke}%
\end{pgfscope}%
\begin{pgfscope}%
\pgfpathrectangle{\pgfqpoint{0.461111in}{0.499691in}}{\pgfqpoint{2.092500in}{2.265000in}} %
\pgfusepath{clip}%
\pgfsetbuttcap%
\pgfsetroundjoin%
\definecolor{currentfill}{rgb}{0.000000,0.000000,0.000000}%
\pgfsetfillcolor{currentfill}%
\pgfsetlinewidth{1.003750pt}%
\definecolor{currentstroke}{rgb}{0.000000,0.000000,0.000000}%
\pgfsetstrokecolor{currentstroke}%
\pgfsetdash{}{0pt}%
\pgfsys@defobject{currentmarker}{\pgfqpoint{-0.041667in}{-0.041667in}}{\pgfqpoint{0.041667in}{0.041667in}}{%
\pgfpathmoveto{\pgfqpoint{0.000000in}{-0.041667in}}%
\pgfpathcurveto{\pgfqpoint{0.011050in}{-0.041667in}}{\pgfqpoint{0.021649in}{-0.037276in}}{\pgfqpoint{0.029463in}{-0.029463in}}%
\pgfpathcurveto{\pgfqpoint{0.037276in}{-0.021649in}}{\pgfqpoint{0.041667in}{-0.011050in}}{\pgfqpoint{0.041667in}{0.000000in}}%
\pgfpathcurveto{\pgfqpoint{0.041667in}{0.011050in}}{\pgfqpoint{0.037276in}{0.021649in}}{\pgfqpoint{0.029463in}{0.029463in}}%
\pgfpathcurveto{\pgfqpoint{0.021649in}{0.037276in}}{\pgfqpoint{0.011050in}{0.041667in}}{\pgfqpoint{0.000000in}{0.041667in}}%
\pgfpathcurveto{\pgfqpoint{-0.011050in}{0.041667in}}{\pgfqpoint{-0.021649in}{0.037276in}}{\pgfqpoint{-0.029463in}{0.029463in}}%
\pgfpathcurveto{\pgfqpoint{-0.037276in}{0.021649in}}{\pgfqpoint{-0.041667in}{0.011050in}}{\pgfqpoint{-0.041667in}{0.000000in}}%
\pgfpathcurveto{\pgfqpoint{-0.041667in}{-0.011050in}}{\pgfqpoint{-0.037276in}{-0.021649in}}{\pgfqpoint{-0.029463in}{-0.029463in}}%
\pgfpathcurveto{\pgfqpoint{-0.021649in}{-0.037276in}}{\pgfqpoint{-0.011050in}{-0.041667in}}{\pgfqpoint{0.000000in}{-0.041667in}}%
\pgfpathclose%
\pgfusepath{stroke,fill}%
}%
\begin{pgfscope}%
\pgfsys@transformshift{0.556225in}{0.602646in}%
\pgfsys@useobject{currentmarker}{}%
\end{pgfscope}%
\begin{pgfscope}%
\pgfsys@transformshift{0.890309in}{1.116868in}%
\pgfsys@useobject{currentmarker}{}%
\end{pgfscope}%
\begin{pgfscope}%
\pgfsys@transformshift{1.137086in}{1.631091in}%
\pgfsys@useobject{currentmarker}{}%
\end{pgfscope}%
\begin{pgfscope}%
\pgfsys@transformshift{1.538610in}{2.659537in}%
\pgfsys@useobject{currentmarker}{}%
\end{pgfscope}%
\end{pgfscope}%
\begin{pgfscope}%
\pgfpathrectangle{\pgfqpoint{0.461111in}{0.499691in}}{\pgfqpoint{2.092500in}{2.265000in}} %
\pgfusepath{clip}%
\pgfsetbuttcap%
\pgfsetroundjoin%
\definecolor{currentfill}{rgb}{1.000000,0.000000,0.000000}%
\pgfsetfillcolor{currentfill}%
\pgfsetlinewidth{1.003750pt}%
\definecolor{currentstroke}{rgb}{1.000000,0.000000,0.000000}%
\pgfsetstrokecolor{currentstroke}%
\pgfsetdash{}{0pt}%
\pgfsys@defobject{currentmarker}{\pgfqpoint{-0.083333in}{-0.083333in}}{\pgfqpoint{0.083333in}{0.083333in}}{%
\pgfpathmoveto{\pgfqpoint{0.000000in}{-0.083333in}}%
\pgfpathcurveto{\pgfqpoint{0.022100in}{-0.083333in}}{\pgfqpoint{0.043298in}{-0.074553in}}{\pgfqpoint{0.058926in}{-0.058926in}}%
\pgfpathcurveto{\pgfqpoint{0.074553in}{-0.043298in}}{\pgfqpoint{0.083333in}{-0.022100in}}{\pgfqpoint{0.083333in}{0.000000in}}%
\pgfpathcurveto{\pgfqpoint{0.083333in}{0.022100in}}{\pgfqpoint{0.074553in}{0.043298in}}{\pgfqpoint{0.058926in}{0.058926in}}%
\pgfpathcurveto{\pgfqpoint{0.043298in}{0.074553in}}{\pgfqpoint{0.022100in}{0.083333in}}{\pgfqpoint{0.000000in}{0.083333in}}%
\pgfpathcurveto{\pgfqpoint{-0.022100in}{0.083333in}}{\pgfqpoint{-0.043298in}{0.074553in}}{\pgfqpoint{-0.058926in}{0.058926in}}%
\pgfpathcurveto{\pgfqpoint{-0.074553in}{0.043298in}}{\pgfqpoint{-0.083333in}{0.022100in}}{\pgfqpoint{-0.083333in}{0.000000in}}%
\pgfpathcurveto{\pgfqpoint{-0.083333in}{-0.022100in}}{\pgfqpoint{-0.074553in}{-0.043298in}}{\pgfqpoint{-0.058926in}{-0.058926in}}%
\pgfpathcurveto{\pgfqpoint{-0.043298in}{-0.074553in}}{\pgfqpoint{-0.022100in}{-0.083333in}}{\pgfqpoint{0.000000in}{-0.083333in}}%
\pgfpathclose%
\pgfusepath{stroke,fill}%
}%
\begin{pgfscope}%
\pgfsys@transformshift{0.556225in}{0.602646in}%
\pgfsys@useobject{currentmarker}{}%
\end{pgfscope}%
\begin{pgfscope}%
\pgfsys@transformshift{1.236484in}{1.116868in}%
\pgfsys@useobject{currentmarker}{}%
\end{pgfscope}%
\begin{pgfscope}%
\pgfsys@transformshift{1.718529in}{1.631091in}%
\pgfsys@useobject{currentmarker}{}%
\end{pgfscope}%
\begin{pgfscope}%
\pgfsys@transformshift{2.458498in}{2.659537in}%
\pgfsys@useobject{currentmarker}{}%
\end{pgfscope}%
\end{pgfscope}%
\begin{pgfscope}%
\pgfpathrectangle{\pgfqpoint{0.461111in}{0.499691in}}{\pgfqpoint{2.092500in}{2.265000in}} %
\pgfusepath{clip}%
\pgfsetbuttcap%
\pgfsetroundjoin%
\pgfsetlinewidth{2.007500pt}%
\definecolor{currentstroke}{rgb}{1.000000,0.000000,0.000000}%
\pgfsetstrokecolor{currentstroke}%
\pgfsetdash{{7.400000pt}{3.200000pt}}{0.000000pt}%
\pgfpathmoveto{\pgfqpoint{0.556225in}{0.602646in}}%
\pgfpathlineto{\pgfqpoint{0.635486in}{0.649932in}}%
\pgfpathlineto{\pgfqpoint{0.714748in}{0.700560in}}%
\pgfpathlineto{\pgfqpoint{0.794009in}{0.754528in}}%
\pgfpathlineto{\pgfqpoint{0.873270in}{0.811838in}}%
\pgfpathlineto{\pgfqpoint{0.952532in}{0.872488in}}%
\pgfpathlineto{\pgfqpoint{1.031793in}{0.936479in}}%
\pgfpathlineto{\pgfqpoint{1.111055in}{1.003811in}}%
\pgfpathlineto{\pgfqpoint{1.190316in}{1.074484in}}%
\pgfpathlineto{\pgfqpoint{1.269577in}{1.148498in}}%
\pgfpathlineto{\pgfqpoint{1.348839in}{1.225853in}}%
\pgfpathlineto{\pgfqpoint{1.428100in}{1.306548in}}%
\pgfpathlineto{\pgfqpoint{1.507361in}{1.390585in}}%
\pgfpathlineto{\pgfqpoint{1.586623in}{1.477962in}}%
\pgfpathlineto{\pgfqpoint{1.665884in}{1.568680in}}%
\pgfpathlineto{\pgfqpoint{1.745145in}{1.662739in}}%
\pgfpathlineto{\pgfqpoint{1.824407in}{1.760139in}}%
\pgfpathlineto{\pgfqpoint{1.903668in}{1.860880in}}%
\pgfpathlineto{\pgfqpoint{1.982930in}{1.964962in}}%
\pgfpathlineto{\pgfqpoint{2.062191in}{2.072385in}}%
\pgfpathlineto{\pgfqpoint{2.141452in}{2.183148in}}%
\pgfpathlineto{\pgfqpoint{2.220714in}{2.297253in}}%
\pgfpathlineto{\pgfqpoint{2.299975in}{2.414698in}}%
\pgfpathlineto{\pgfqpoint{2.379236in}{2.535485in}}%
\pgfpathlineto{\pgfqpoint{2.458498in}{2.659612in}}%
\pgfusepath{stroke}%
\end{pgfscope}%
\begin{pgfscope}%
\pgfpathrectangle{\pgfqpoint{0.461111in}{0.499691in}}{\pgfqpoint{2.092500in}{2.265000in}} %
\pgfusepath{clip}%
\pgfsetbuttcap%
\pgfsetmiterjoin%
\definecolor{currentfill}{rgb}{0.000000,0.500000,0.000000}%
\pgfsetfillcolor{currentfill}%
\pgfsetlinewidth{1.003750pt}%
\definecolor{currentstroke}{rgb}{0.000000,0.500000,0.000000}%
\pgfsetstrokecolor{currentstroke}%
\pgfsetdash{}{0pt}%
\pgfsys@defobject{currentmarker}{\pgfqpoint{-0.083333in}{-0.083333in}}{\pgfqpoint{0.083333in}{0.083333in}}{%
\pgfpathmoveto{\pgfqpoint{-0.083333in}{-0.083333in}}%
\pgfpathlineto{\pgfqpoint{0.083333in}{-0.083333in}}%
\pgfpathlineto{\pgfqpoint{0.083333in}{0.083333in}}%
\pgfpathlineto{\pgfqpoint{-0.083333in}{0.083333in}}%
\pgfpathclose%
\pgfusepath{stroke,fill}%
}%
\begin{pgfscope}%
\pgfsys@transformshift{0.556225in}{0.602646in}%
\pgfsys@useobject{currentmarker}{}%
\end{pgfscope}%
\begin{pgfscope}%
\pgfsys@transformshift{0.771236in}{1.116868in}%
\pgfsys@useobject{currentmarker}{}%
\end{pgfscope}%
\begin{pgfscope}%
\pgfsys@transformshift{0.940912in}{1.631091in}%
\pgfsys@useobject{currentmarker}{}%
\end{pgfscope}%
\begin{pgfscope}%
\pgfsys@transformshift{1.217587in}{2.659537in}%
\pgfsys@useobject{currentmarker}{}%
\end{pgfscope}%
\end{pgfscope}%
\begin{pgfscope}%
\pgfpathrectangle{\pgfqpoint{0.461111in}{0.499691in}}{\pgfqpoint{2.092500in}{2.265000in}} %
\pgfusepath{clip}%
\pgfsetbuttcap%
\pgfsetroundjoin%
\pgfsetlinewidth{2.007500pt}%
\definecolor{currentstroke}{rgb}{0.000000,0.500000,0.000000}%
\pgfsetstrokecolor{currentstroke}%
\pgfsetdash{{7.400000pt}{3.200000pt}}{0.000000pt}%
\pgfpathmoveto{\pgfqpoint{0.556225in}{0.602646in}}%
\pgfpathlineto{\pgfqpoint{0.583782in}{0.660528in}}%
\pgfpathlineto{\pgfqpoint{0.611339in}{0.720830in}}%
\pgfpathlineto{\pgfqpoint{0.638895in}{0.783554in}}%
\pgfpathlineto{\pgfqpoint{0.666452in}{0.848697in}}%
\pgfpathlineto{\pgfqpoint{0.694009in}{0.916261in}}%
\pgfpathlineto{\pgfqpoint{0.721566in}{0.986245in}}%
\pgfpathlineto{\pgfqpoint{0.749122in}{1.058650in}}%
\pgfpathlineto{\pgfqpoint{0.776679in}{1.133476in}}%
\pgfpathlineto{\pgfqpoint{0.804236in}{1.210721in}}%
\pgfpathlineto{\pgfqpoint{0.831793in}{1.290388in}}%
\pgfpathlineto{\pgfqpoint{0.859349in}{1.372474in}}%
\pgfpathlineto{\pgfqpoint{0.886906in}{1.456982in}}%
\pgfpathlineto{\pgfqpoint{0.914463in}{1.543909in}}%
\pgfpathlineto{\pgfqpoint{0.942020in}{1.633257in}}%
\pgfpathlineto{\pgfqpoint{0.969577in}{1.725026in}}%
\pgfpathlineto{\pgfqpoint{0.997133in}{1.819215in}}%
\pgfpathlineto{\pgfqpoint{1.024690in}{1.915824in}}%
\pgfpathlineto{\pgfqpoint{1.052247in}{2.014854in}}%
\pgfpathlineto{\pgfqpoint{1.079804in}{2.116304in}}%
\pgfpathlineto{\pgfqpoint{1.107360in}{2.220175in}}%
\pgfpathlineto{\pgfqpoint{1.134917in}{2.326466in}}%
\pgfpathlineto{\pgfqpoint{1.162474in}{2.435178in}}%
\pgfpathlineto{\pgfqpoint{1.190031in}{2.546310in}}%
\pgfpathlineto{\pgfqpoint{1.217587in}{2.659863in}}%
\pgfusepath{stroke}%
\end{pgfscope}%
\begin{pgfscope}%
\pgfpathrectangle{\pgfqpoint{0.461111in}{0.499691in}}{\pgfqpoint{2.092500in}{2.265000in}} %
\pgfusepath{clip}%
\pgfsetbuttcap%
\pgfsetbeveljoin%
\definecolor{currentfill}{rgb}{0.000000,0.000000,1.000000}%
\pgfsetfillcolor{currentfill}%
\pgfsetlinewidth{1.003750pt}%
\definecolor{currentstroke}{rgb}{0.000000,0.000000,1.000000}%
\pgfsetstrokecolor{currentstroke}%
\pgfsetdash{}{0pt}%
\pgfsys@defobject{currentmarker}{\pgfqpoint{-0.079255in}{-0.067418in}}{\pgfqpoint{0.079255in}{0.083333in}}{%
\pgfpathmoveto{\pgfqpoint{0.000000in}{0.083333in}}%
\pgfpathlineto{\pgfqpoint{-0.018709in}{0.025751in}}%
\pgfpathlineto{\pgfqpoint{-0.079255in}{0.025751in}}%
\pgfpathlineto{\pgfqpoint{-0.030273in}{-0.009836in}}%
\pgfpathlineto{\pgfqpoint{-0.048982in}{-0.067418in}}%
\pgfpathlineto{\pgfqpoint{-0.000000in}{-0.031831in}}%
\pgfpathlineto{\pgfqpoint{0.048982in}{-0.067418in}}%
\pgfpathlineto{\pgfqpoint{0.030273in}{-0.009836in}}%
\pgfpathlineto{\pgfqpoint{0.079255in}{0.025751in}}%
\pgfpathlineto{\pgfqpoint{0.018709in}{0.025751in}}%
\pgfpathclose%
\pgfusepath{stroke,fill}%
}%
\begin{pgfscope}%
\pgfsys@transformshift{0.556225in}{0.602646in}%
\pgfsys@useobject{currentmarker}{}%
\end{pgfscope}%
\begin{pgfscope}%
\pgfsys@transformshift{0.722898in}{1.116868in}%
\pgfsys@useobject{currentmarker}{}%
\end{pgfscope}%
\begin{pgfscope}%
\pgfsys@transformshift{0.863360in}{1.631091in}%
\pgfsys@useobject{currentmarker}{}%
\end{pgfscope}%
\begin{pgfscope}%
\pgfsys@transformshift{1.095266in}{2.659537in}%
\pgfsys@useobject{currentmarker}{}%
\end{pgfscope}%
\end{pgfscope}%
\begin{pgfscope}%
\pgfpathrectangle{\pgfqpoint{0.461111in}{0.499691in}}{\pgfqpoint{2.092500in}{2.265000in}} %
\pgfusepath{clip}%
\pgfsetbuttcap%
\pgfsetroundjoin%
\pgfsetlinewidth{2.007500pt}%
\definecolor{currentstroke}{rgb}{0.000000,0.000000,1.000000}%
\pgfsetstrokecolor{currentstroke}%
\pgfsetdash{{7.400000pt}{3.200000pt}}{0.000000pt}%
\pgfpathmoveto{\pgfqpoint{0.556225in}{0.602646in}}%
\pgfpathlineto{\pgfqpoint{0.578685in}{0.665247in}}%
\pgfpathlineto{\pgfqpoint{0.601145in}{0.729857in}}%
\pgfpathlineto{\pgfqpoint{0.623605in}{0.796473in}}%
\pgfpathlineto{\pgfqpoint{0.646065in}{0.865098in}}%
\pgfpathlineto{\pgfqpoint{0.668525in}{0.935730in}}%
\pgfpathlineto{\pgfqpoint{0.690985in}{1.008370in}}%
\pgfpathlineto{\pgfqpoint{0.713445in}{1.083018in}}%
\pgfpathlineto{\pgfqpoint{0.735905in}{1.159673in}}%
\pgfpathlineto{\pgfqpoint{0.758366in}{1.238335in}}%
\pgfpathlineto{\pgfqpoint{0.780826in}{1.319006in}}%
\pgfpathlineto{\pgfqpoint{0.803286in}{1.401684in}}%
\pgfpathlineto{\pgfqpoint{0.825746in}{1.486369in}}%
\pgfpathlineto{\pgfqpoint{0.848206in}{1.573063in}}%
\pgfpathlineto{\pgfqpoint{0.870666in}{1.661764in}}%
\pgfpathlineto{\pgfqpoint{0.893126in}{1.752472in}}%
\pgfpathlineto{\pgfqpoint{0.915586in}{1.845188in}}%
\pgfpathlineto{\pgfqpoint{0.938046in}{1.939912in}}%
\pgfpathlineto{\pgfqpoint{0.960506in}{2.036644in}}%
\pgfpathlineto{\pgfqpoint{0.982966in}{2.135383in}}%
\pgfpathlineto{\pgfqpoint{1.005426in}{2.236130in}}%
\pgfpathlineto{\pgfqpoint{1.027886in}{2.338884in}}%
\pgfpathlineto{\pgfqpoint{1.050346in}{2.443646in}}%
\pgfpathlineto{\pgfqpoint{1.072806in}{2.550416in}}%
\pgfpathlineto{\pgfqpoint{1.095266in}{2.659193in}}%
\pgfusepath{stroke}%
\end{pgfscope}%
\begin{pgfscope}%
\pgfpathrectangle{\pgfqpoint{0.461111in}{0.499691in}}{\pgfqpoint{2.092500in}{2.265000in}} %
\pgfusepath{clip}%
\pgfsetbuttcap%
\pgfsetmiterjoin%
\definecolor{currentfill}{rgb}{0.000000,0.750000,0.750000}%
\pgfsetfillcolor{currentfill}%
\pgfsetlinewidth{1.003750pt}%
\definecolor{currentstroke}{rgb}{0.000000,0.750000,0.750000}%
\pgfsetstrokecolor{currentstroke}%
\pgfsetdash{}{0pt}%
\pgfsys@defobject{currentmarker}{\pgfqpoint{-0.083333in}{-0.083333in}}{\pgfqpoint{0.083333in}{0.083333in}}{%
\pgfpathmoveto{\pgfqpoint{0.000000in}{0.083333in}}%
\pgfpathlineto{\pgfqpoint{-0.083333in}{-0.083333in}}%
\pgfpathlineto{\pgfqpoint{0.083333in}{-0.083333in}}%
\pgfpathclose%
\pgfusepath{stroke,fill}%
}%
\begin{pgfscope}%
\pgfsys@transformshift{0.556225in}{0.602646in}%
\pgfsys@useobject{currentmarker}{}%
\end{pgfscope}%
\begin{pgfscope}%
\pgfsys@transformshift{0.556374in}{1.116868in}%
\pgfsys@useobject{currentmarker}{}%
\end{pgfscope}%
\begin{pgfscope}%
\pgfsys@transformshift{0.556386in}{1.631091in}%
\pgfsys@useobject{currentmarker}{}%
\end{pgfscope}%
\begin{pgfscope}%
\pgfsys@transformshift{0.556751in}{2.659537in}%
\pgfsys@useobject{currentmarker}{}%
\end{pgfscope}%
\end{pgfscope}%
\begin{pgfscope}%
\pgfpathrectangle{\pgfqpoint{0.461111in}{0.499691in}}{\pgfqpoint{2.092500in}{2.265000in}} %
\pgfusepath{clip}%
\pgfsetbuttcap%
\pgfsetroundjoin%
\pgfsetlinewidth{2.007500pt}%
\definecolor{currentstroke}{rgb}{0.000000,0.750000,0.750000}%
\pgfsetstrokecolor{currentstroke}%
\pgfsetdash{{7.400000pt}{3.200000pt}}{0.000000pt}%
\pgfpathmoveto{\pgfqpoint{0.556225in}{0.602646in}}%
\pgfpathlineto{\pgfqpoint{0.556247in}{0.721432in}}%
\pgfpathlineto{\pgfqpoint{0.556269in}{0.837350in}}%
\pgfpathlineto{\pgfqpoint{0.556291in}{0.950400in}}%
\pgfpathlineto{\pgfqpoint{0.556313in}{1.060580in}}%
\pgfpathlineto{\pgfqpoint{0.556335in}{1.167891in}}%
\pgfpathlineto{\pgfqpoint{0.556357in}{1.272334in}}%
\pgfpathlineto{\pgfqpoint{0.556379in}{1.373908in}}%
\pgfpathlineto{\pgfqpoint{0.556400in}{1.472613in}}%
\pgfpathlineto{\pgfqpoint{0.556422in}{1.568450in}}%
\pgfpathlineto{\pgfqpoint{0.556444in}{1.661417in}}%
\pgfpathlineto{\pgfqpoint{0.556466in}{1.751516in}}%
\pgfpathlineto{\pgfqpoint{0.556488in}{1.838746in}}%
\pgfpathlineto{\pgfqpoint{0.556510in}{1.923107in}}%
\pgfpathlineto{\pgfqpoint{0.556532in}{2.004599in}}%
\pgfpathlineto{\pgfqpoint{0.556554in}{2.083222in}}%
\pgfpathlineto{\pgfqpoint{0.556576in}{2.158977in}}%
\pgfpathlineto{\pgfqpoint{0.556598in}{2.231863in}}%
\pgfpathlineto{\pgfqpoint{0.556620in}{2.301880in}}%
\pgfpathlineto{\pgfqpoint{0.556642in}{2.369028in}}%
\pgfpathlineto{\pgfqpoint{0.556664in}{2.433307in}}%
\pgfpathlineto{\pgfqpoint{0.556686in}{2.494718in}}%
\pgfpathlineto{\pgfqpoint{0.556708in}{2.553260in}}%
\pgfpathlineto{\pgfqpoint{0.556730in}{2.608932in}}%
\pgfpathlineto{\pgfqpoint{0.556751in}{2.661737in}}%
\pgfusepath{stroke}%
\end{pgfscope}%
\begin{pgfscope}%
\pgfsetrectcap%
\pgfsetmiterjoin%
\pgfsetlinewidth{0.803000pt}%
\definecolor{currentstroke}{rgb}{0.000000,0.000000,0.000000}%
\pgfsetstrokecolor{currentstroke}%
\pgfsetdash{}{0pt}%
\pgfpathmoveto{\pgfqpoint{0.461111in}{0.499691in}}%
\pgfpathlineto{\pgfqpoint{0.461111in}{2.764691in}}%
\pgfusepath{stroke}%
\end{pgfscope}%
\begin{pgfscope}%
\pgfsetrectcap%
\pgfsetmiterjoin%
\pgfsetlinewidth{0.803000pt}%
\definecolor{currentstroke}{rgb}{0.000000,0.000000,0.000000}%
\pgfsetstrokecolor{currentstroke}%
\pgfsetdash{}{0pt}%
\pgfpathmoveto{\pgfqpoint{2.553611in}{0.499691in}}%
\pgfpathlineto{\pgfqpoint{2.553611in}{2.764691in}}%
\pgfusepath{stroke}%
\end{pgfscope}%
\begin{pgfscope}%
\pgfsetrectcap%
\pgfsetmiterjoin%
\pgfsetlinewidth{0.803000pt}%
\definecolor{currentstroke}{rgb}{0.000000,0.000000,0.000000}%
\pgfsetstrokecolor{currentstroke}%
\pgfsetdash{}{0pt}%
\pgfpathmoveto{\pgfqpoint{0.461111in}{0.499691in}}%
\pgfpathlineto{\pgfqpoint{2.553611in}{0.499691in}}%
\pgfusepath{stroke}%
\end{pgfscope}%
\begin{pgfscope}%
\pgfsetrectcap%
\pgfsetmiterjoin%
\pgfsetlinewidth{0.803000pt}%
\definecolor{currentstroke}{rgb}{0.000000,0.000000,0.000000}%
\pgfsetstrokecolor{currentstroke}%
\pgfsetdash{}{0pt}%
\pgfpathmoveto{\pgfqpoint{0.461111in}{2.764691in}}%
\pgfpathlineto{\pgfqpoint{2.553611in}{2.764691in}}%
\pgfusepath{stroke}%
\end{pgfscope}%
\begin{pgfscope}%
\pgfsetbuttcap%
\pgfsetmiterjoin%
\definecolor{currentfill}{rgb}{1.000000,1.000000,1.000000}%
\pgfsetfillcolor{currentfill}%
\pgfsetfillopacity{0.800000}%
\pgfsetlinewidth{1.003750pt}%
\definecolor{currentstroke}{rgb}{0.800000,0.800000,0.800000}%
\pgfsetstrokecolor{currentstroke}%
\pgfsetstrokeopacity{0.800000}%
\pgfsetdash{}{0pt}%
\pgfpathmoveto{\pgfqpoint{1.853380in}{0.569136in}}%
\pgfpathlineto{\pgfqpoint{2.456389in}{0.569136in}}%
\pgfpathquadraticcurveto{\pgfqpoint{2.484167in}{0.569136in}}{\pgfqpoint{2.484167in}{0.596913in}}%
\pgfpathlineto{\pgfqpoint{2.484167in}{1.551388in}}%
\pgfpathquadraticcurveto{\pgfqpoint{2.484167in}{1.579166in}}{\pgfqpoint{2.456389in}{1.579166in}}%
\pgfpathlineto{\pgfqpoint{1.853380in}{1.579166in}}%
\pgfpathquadraticcurveto{\pgfqpoint{1.825602in}{1.579166in}}{\pgfqpoint{1.825602in}{1.551388in}}%
\pgfpathlineto{\pgfqpoint{1.825602in}{0.596913in}}%
\pgfpathquadraticcurveto{\pgfqpoint{1.825602in}{0.569136in}}{\pgfqpoint{1.853380in}{0.569136in}}%
\pgfpathclose%
\pgfusepath{stroke,fill}%
\end{pgfscope}%
\begin{pgfscope}%
\pgfsetbuttcap%
\pgfsetroundjoin%
\pgfsetlinewidth{1.505625pt}%
\definecolor{currentstroke}{rgb}{1.000000,0.000000,0.000000}%
\pgfsetstrokecolor{currentstroke}%
\pgfsetdash{{5.550000pt}{2.400000pt}}{0.000000pt}%
\pgfpathmoveto{\pgfqpoint{1.881157in}{1.474999in}}%
\pgfpathlineto{\pgfqpoint{2.158935in}{1.474999in}}%
\pgfusepath{stroke}%
\end{pgfscope}%
\begin{pgfscope}%
\pgfsetbuttcap%
\pgfsetroundjoin%
\definecolor{currentfill}{rgb}{1.000000,0.000000,0.000000}%
\pgfsetfillcolor{currentfill}%
\pgfsetlinewidth{1.003750pt}%
\definecolor{currentstroke}{rgb}{1.000000,0.000000,0.000000}%
\pgfsetstrokecolor{currentstroke}%
\pgfsetdash{}{0pt}%
\pgfsys@defobject{currentmarker}{\pgfqpoint{-0.041667in}{-0.041667in}}{\pgfqpoint{0.041667in}{0.041667in}}{%
\pgfpathmoveto{\pgfqpoint{0.000000in}{-0.041667in}}%
\pgfpathcurveto{\pgfqpoint{0.011050in}{-0.041667in}}{\pgfqpoint{0.021649in}{-0.037276in}}{\pgfqpoint{0.029463in}{-0.029463in}}%
\pgfpathcurveto{\pgfqpoint{0.037276in}{-0.021649in}}{\pgfqpoint{0.041667in}{-0.011050in}}{\pgfqpoint{0.041667in}{0.000000in}}%
\pgfpathcurveto{\pgfqpoint{0.041667in}{0.011050in}}{\pgfqpoint{0.037276in}{0.021649in}}{\pgfqpoint{0.029463in}{0.029463in}}%
\pgfpathcurveto{\pgfqpoint{0.021649in}{0.037276in}}{\pgfqpoint{0.011050in}{0.041667in}}{\pgfqpoint{0.000000in}{0.041667in}}%
\pgfpathcurveto{\pgfqpoint{-0.011050in}{0.041667in}}{\pgfqpoint{-0.021649in}{0.037276in}}{\pgfqpoint{-0.029463in}{0.029463in}}%
\pgfpathcurveto{\pgfqpoint{-0.037276in}{0.021649in}}{\pgfqpoint{-0.041667in}{0.011050in}}{\pgfqpoint{-0.041667in}{0.000000in}}%
\pgfpathcurveto{\pgfqpoint{-0.041667in}{-0.011050in}}{\pgfqpoint{-0.037276in}{-0.021649in}}{\pgfqpoint{-0.029463in}{-0.029463in}}%
\pgfpathcurveto{\pgfqpoint{-0.021649in}{-0.037276in}}{\pgfqpoint{-0.011050in}{-0.041667in}}{\pgfqpoint{0.000000in}{-0.041667in}}%
\pgfpathclose%
\pgfusepath{stroke,fill}%
}%
\begin{pgfscope}%
\pgfsys@transformshift{2.020046in}{1.474999in}%
\pgfsys@useobject{currentmarker}{}%
\end{pgfscope}%
\end{pgfscope}%
\begin{pgfscope}%
\pgftext[x=2.270046in,y=1.426388in,left,base]{\rmfamily\fontsize{10.000000}{12.000000}\selectfont A}%
\end{pgfscope}%
\begin{pgfscope}%
\pgfsetbuttcap%
\pgfsetroundjoin%
\pgfsetlinewidth{1.505625pt}%
\definecolor{currentstroke}{rgb}{0.000000,0.501961,0.000000}%
\pgfsetstrokecolor{currentstroke}%
\pgfsetdash{{5.550000pt}{2.400000pt}}{0.000000pt}%
\pgfpathmoveto{\pgfqpoint{1.881157in}{1.281327in}}%
\pgfpathlineto{\pgfqpoint{2.158935in}{1.281327in}}%
\pgfusepath{stroke}%
\end{pgfscope}%
\begin{pgfscope}%
\pgfsetbuttcap%
\pgfsetmiterjoin%
\definecolor{currentfill}{rgb}{0.000000,0.501961,0.000000}%
\pgfsetfillcolor{currentfill}%
\pgfsetlinewidth{1.003750pt}%
\definecolor{currentstroke}{rgb}{0.000000,0.501961,0.000000}%
\pgfsetstrokecolor{currentstroke}%
\pgfsetdash{}{0pt}%
\pgfsys@defobject{currentmarker}{\pgfqpoint{-0.041667in}{-0.041667in}}{\pgfqpoint{0.041667in}{0.041667in}}{%
\pgfpathmoveto{\pgfqpoint{-0.041667in}{-0.041667in}}%
\pgfpathlineto{\pgfqpoint{0.041667in}{-0.041667in}}%
\pgfpathlineto{\pgfqpoint{0.041667in}{0.041667in}}%
\pgfpathlineto{\pgfqpoint{-0.041667in}{0.041667in}}%
\pgfpathclose%
\pgfusepath{stroke,fill}%
}%
\begin{pgfscope}%
\pgfsys@transformshift{2.020046in}{1.281327in}%
\pgfsys@useobject{currentmarker}{}%
\end{pgfscope}%
\end{pgfscope}%
\begin{pgfscope}%
\pgftext[x=2.270046in,y=1.232716in,left,base]{\rmfamily\fontsize{10.000000}{12.000000}\selectfont B}%
\end{pgfscope}%
\begin{pgfscope}%
\pgfsetbuttcap%
\pgfsetroundjoin%
\pgfsetlinewidth{1.505625pt}%
\definecolor{currentstroke}{rgb}{0.000000,0.000000,1.000000}%
\pgfsetstrokecolor{currentstroke}%
\pgfsetdash{{5.550000pt}{2.400000pt}}{0.000000pt}%
\pgfpathmoveto{\pgfqpoint{1.881157in}{1.087654in}}%
\pgfpathlineto{\pgfqpoint{2.158935in}{1.087654in}}%
\pgfusepath{stroke}%
\end{pgfscope}%
\begin{pgfscope}%
\pgfsetbuttcap%
\pgfsetbeveljoin%
\definecolor{currentfill}{rgb}{0.000000,0.000000,1.000000}%
\pgfsetfillcolor{currentfill}%
\pgfsetlinewidth{1.003750pt}%
\definecolor{currentstroke}{rgb}{0.000000,0.000000,1.000000}%
\pgfsetstrokecolor{currentstroke}%
\pgfsetdash{}{0pt}%
\pgfsys@defobject{currentmarker}{\pgfqpoint{-0.039627in}{-0.033709in}}{\pgfqpoint{0.039627in}{0.041667in}}{%
\pgfpathmoveto{\pgfqpoint{0.000000in}{0.041667in}}%
\pgfpathlineto{\pgfqpoint{-0.009355in}{0.012876in}}%
\pgfpathlineto{\pgfqpoint{-0.039627in}{0.012876in}}%
\pgfpathlineto{\pgfqpoint{-0.015136in}{-0.004918in}}%
\pgfpathlineto{\pgfqpoint{-0.024491in}{-0.033709in}}%
\pgfpathlineto{\pgfqpoint{-0.000000in}{-0.015915in}}%
\pgfpathlineto{\pgfqpoint{0.024491in}{-0.033709in}}%
\pgfpathlineto{\pgfqpoint{0.015136in}{-0.004918in}}%
\pgfpathlineto{\pgfqpoint{0.039627in}{0.012876in}}%
\pgfpathlineto{\pgfqpoint{0.009355in}{0.012876in}}%
\pgfpathclose%
\pgfusepath{stroke,fill}%
}%
\begin{pgfscope}%
\pgfsys@transformshift{2.020046in}{1.087654in}%
\pgfsys@useobject{currentmarker}{}%
\end{pgfscope}%
\end{pgfscope}%
\begin{pgfscope}%
\pgftext[x=2.270046in,y=1.039043in,left,base]{\rmfamily\fontsize{10.000000}{12.000000}\selectfont C}%
\end{pgfscope}%
\begin{pgfscope}%
\pgfsetbuttcap%
\pgfsetroundjoin%
\pgfsetlinewidth{1.505625pt}%
\definecolor{currentstroke}{rgb}{0.000000,0.750000,0.750000}%
\pgfsetstrokecolor{currentstroke}%
\pgfsetdash{{5.550000pt}{2.400000pt}}{0.000000pt}%
\pgfpathmoveto{\pgfqpoint{1.881157in}{0.893981in}}%
\pgfpathlineto{\pgfqpoint{2.158935in}{0.893981in}}%
\pgfusepath{stroke}%
\end{pgfscope}%
\begin{pgfscope}%
\pgfsetbuttcap%
\pgfsetmiterjoin%
\definecolor{currentfill}{rgb}{0.000000,0.750000,0.750000}%
\pgfsetfillcolor{currentfill}%
\pgfsetlinewidth{1.003750pt}%
\definecolor{currentstroke}{rgb}{0.000000,0.750000,0.750000}%
\pgfsetstrokecolor{currentstroke}%
\pgfsetdash{}{0pt}%
\pgfsys@defobject{currentmarker}{\pgfqpoint{-0.041667in}{-0.041667in}}{\pgfqpoint{0.041667in}{0.041667in}}{%
\pgfpathmoveto{\pgfqpoint{0.000000in}{0.041667in}}%
\pgfpathlineto{\pgfqpoint{-0.041667in}{-0.041667in}}%
\pgfpathlineto{\pgfqpoint{0.041667in}{-0.041667in}}%
\pgfpathclose%
\pgfusepath{stroke,fill}%
}%
\begin{pgfscope}%
\pgfsys@transformshift{2.020046in}{0.893981in}%
\pgfsys@useobject{currentmarker}{}%
\end{pgfscope}%
\end{pgfscope}%
\begin{pgfscope}%
\pgftext[x=2.270046in,y=0.845370in,left,base]{\rmfamily\fontsize{10.000000}{12.000000}\selectfont D}%
\end{pgfscope}%
\begin{pgfscope}%
\pgfsetrectcap%
\pgfsetroundjoin%
\pgfsetlinewidth{1.505625pt}%
\definecolor{currentstroke}{rgb}{0.000000,0.000000,0.000000}%
\pgfsetstrokecolor{currentstroke}%
\pgfsetdash{}{0pt}%
\pgfpathmoveto{\pgfqpoint{1.881157in}{0.700308in}}%
\pgfpathlineto{\pgfqpoint{2.158935in}{0.700308in}}%
\pgfusepath{stroke}%
\end{pgfscope}%
\begin{pgfscope}%
\pgfsetbuttcap%
\pgfsetroundjoin%
\definecolor{currentfill}{rgb}{0.000000,0.000000,0.000000}%
\pgfsetfillcolor{currentfill}%
\pgfsetlinewidth{1.003750pt}%
\definecolor{currentstroke}{rgb}{0.000000,0.000000,0.000000}%
\pgfsetstrokecolor{currentstroke}%
\pgfsetdash{}{0pt}%
\pgfsys@defobject{currentmarker}{\pgfqpoint{-0.020833in}{-0.020833in}}{\pgfqpoint{0.020833in}{0.020833in}}{%
\pgfpathmoveto{\pgfqpoint{0.000000in}{-0.020833in}}%
\pgfpathcurveto{\pgfqpoint{0.005525in}{-0.020833in}}{\pgfqpoint{0.010825in}{-0.018638in}}{\pgfqpoint{0.014731in}{-0.014731in}}%
\pgfpathcurveto{\pgfqpoint{0.018638in}{-0.010825in}}{\pgfqpoint{0.020833in}{-0.005525in}}{\pgfqpoint{0.020833in}{0.000000in}}%
\pgfpathcurveto{\pgfqpoint{0.020833in}{0.005525in}}{\pgfqpoint{0.018638in}{0.010825in}}{\pgfqpoint{0.014731in}{0.014731in}}%
\pgfpathcurveto{\pgfqpoint{0.010825in}{0.018638in}}{\pgfqpoint{0.005525in}{0.020833in}}{\pgfqpoint{0.000000in}{0.020833in}}%
\pgfpathcurveto{\pgfqpoint{-0.005525in}{0.020833in}}{\pgfqpoint{-0.010825in}{0.018638in}}{\pgfqpoint{-0.014731in}{0.014731in}}%
\pgfpathcurveto{\pgfqpoint{-0.018638in}{0.010825in}}{\pgfqpoint{-0.020833in}{0.005525in}}{\pgfqpoint{-0.020833in}{0.000000in}}%
\pgfpathcurveto{\pgfqpoint{-0.020833in}{-0.005525in}}{\pgfqpoint{-0.018638in}{-0.010825in}}{\pgfqpoint{-0.014731in}{-0.014731in}}%
\pgfpathcurveto{\pgfqpoint{-0.010825in}{-0.018638in}}{\pgfqpoint{-0.005525in}{-0.020833in}}{\pgfqpoint{0.000000in}{-0.020833in}}%
\pgfpathclose%
\pgfusepath{stroke,fill}%
}%
\begin{pgfscope}%
\pgfsys@transformshift{2.020046in}{0.700308in}%
\pgfsys@useobject{currentmarker}{}%
\end{pgfscope}%
\end{pgfscope}%
\begin{pgfscope}%
\pgftext[x=2.270046in,y=0.651697in,left,base]{\rmfamily\fontsize{10.000000}{12.000000}\selectfont ref}%
\end{pgfscope}%
\end{pgfpicture}%
\makeatother%
\endgroup%

%% file: convergence.pgf
\begingroup%
\makeatletter%
\begin{pgfpicture}%
\pgfpathrectangle{\pgfpointorigin}{\pgfqpoint{2.895499in}{2.876358in}}%
\pgfusepath{use as bounding box, clip}%
\begin{pgfscope}%
\pgfsetbuttcap%
\pgfsetmiterjoin%
\definecolor{currentfill}{rgb}{1.000000,1.000000,1.000000}%
\pgfsetfillcolor{currentfill}%
\pgfsetlinewidth{0.000000pt}%
\definecolor{currentstroke}{rgb}{1.000000,1.000000,1.000000}%
\pgfsetstrokecolor{currentstroke}%
\pgfsetdash{}{0pt}%
\pgfpathmoveto{\pgfqpoint{0.000000in}{0.000000in}}%
\pgfpathlineto{\pgfqpoint{2.895499in}{0.000000in}}%
\pgfpathlineto{\pgfqpoint{2.895499in}{2.876358in}}%
\pgfpathlineto{\pgfqpoint{0.000000in}{2.876358in}}%
\pgfpathclose%
\pgfusepath{fill}%
\end{pgfscope}%
\begin{pgfscope}%
\pgfsetbuttcap%
\pgfsetmiterjoin%
\definecolor{currentfill}{rgb}{1.000000,1.000000,1.000000}%
\pgfsetfillcolor{currentfill}%
\pgfsetlinewidth{0.000000pt}%
\definecolor{currentstroke}{rgb}{0.000000,0.000000,0.000000}%
\pgfsetstrokecolor{currentstroke}%
\pgfsetstrokeopacity{0.000000}%
\pgfsetdash{}{0pt}%
\pgfpathmoveto{\pgfqpoint{0.638581in}{0.499691in}}%
\pgfpathlineto{\pgfqpoint{2.731081in}{0.499691in}}%
\pgfpathlineto{\pgfqpoint{2.731081in}{2.764691in}}%
\pgfpathlineto{\pgfqpoint{0.638581in}{2.764691in}}%
\pgfpathclose%
\pgfusepath{fill}%
\end{pgfscope}%
\begin{pgfscope}%
\pgfsetbuttcap%
\pgfsetroundjoin%
\definecolor{currentfill}{rgb}{0.000000,0.000000,0.000000}%
\pgfsetfillcolor{currentfill}%
\pgfsetlinewidth{0.803000pt}%
\definecolor{currentstroke}{rgb}{0.000000,0.000000,0.000000}%
\pgfsetstrokecolor{currentstroke}%
\pgfsetdash{}{0pt}%
\pgfsys@defobject{currentmarker}{\pgfqpoint{0.000000in}{-0.048611in}}{\pgfqpoint{0.000000in}{0.000000in}}{%
\pgfpathmoveto{\pgfqpoint{0.000000in}{0.000000in}}%
\pgfpathlineto{\pgfqpoint{0.000000in}{-0.048611in}}%
\pgfusepath{stroke,fill}%
}%
\begin{pgfscope}%
\pgfsys@transformshift{0.763164in}{0.499691in}%
\pgfsys@useobject{currentmarker}{}%
\end{pgfscope}%
\end{pgfscope}%
\begin{pgfscope}%
\definecolor{textcolor}{rgb}{0.000000,0.000000,0.000000}%
\pgfsetstrokecolor{textcolor}%
\pgfsetfillcolor{textcolor}%
\pgftext[x=0.763164in,y=0.402469in,,top]{\color{textcolor}\rmfamily\fontsize{10.000000}{12.000000}\selectfont \(\displaystyle 10^{2}\)}%
\end{pgfscope}%
\begin{pgfscope}%
\pgfsetbuttcap%
\pgfsetroundjoin%
\definecolor{currentfill}{rgb}{0.000000,0.000000,0.000000}%
\pgfsetfillcolor{currentfill}%
\pgfsetlinewidth{0.803000pt}%
\definecolor{currentstroke}{rgb}{0.000000,0.000000,0.000000}%
\pgfsetstrokecolor{currentstroke}%
\pgfsetdash{}{0pt}%
\pgfsys@defobject{currentmarker}{\pgfqpoint{0.000000in}{-0.048611in}}{\pgfqpoint{0.000000in}{0.000000in}}{%
\pgfpathmoveto{\pgfqpoint{0.000000in}{0.000000in}}%
\pgfpathlineto{\pgfqpoint{0.000000in}{-0.048611in}}%
\pgfusepath{stroke,fill}%
}%
\begin{pgfscope}%
\pgfsys@transformshift{1.407076in}{0.499691in}%
\pgfsys@useobject{currentmarker}{}%
\end{pgfscope}%
\end{pgfscope}%
\begin{pgfscope}%
\definecolor{textcolor}{rgb}{0.000000,0.000000,0.000000}%
\pgfsetstrokecolor{textcolor}%
\pgfsetfillcolor{textcolor}%
\pgftext[x=1.407076in,y=0.402469in,,top]{\color{textcolor}\rmfamily\fontsize{10.000000}{12.000000}\selectfont \(\displaystyle 10^{3}\)}%
\end{pgfscope}%
\begin{pgfscope}%
\pgfsetbuttcap%
\pgfsetroundjoin%
\definecolor{currentfill}{rgb}{0.000000,0.000000,0.000000}%
\pgfsetfillcolor{currentfill}%
\pgfsetlinewidth{0.803000pt}%
\definecolor{currentstroke}{rgb}{0.000000,0.000000,0.000000}%
\pgfsetstrokecolor{currentstroke}%
\pgfsetdash{}{0pt}%
\pgfsys@defobject{currentmarker}{\pgfqpoint{0.000000in}{-0.048611in}}{\pgfqpoint{0.000000in}{0.000000in}}{%
\pgfpathmoveto{\pgfqpoint{0.000000in}{0.000000in}}%
\pgfpathlineto{\pgfqpoint{0.000000in}{-0.048611in}}%
\pgfusepath{stroke,fill}%
}%
\begin{pgfscope}%
\pgfsys@transformshift{2.050988in}{0.499691in}%
\pgfsys@useobject{currentmarker}{}%
\end{pgfscope}%
\end{pgfscope}%
\begin{pgfscope}%
\definecolor{textcolor}{rgb}{0.000000,0.000000,0.000000}%
\pgfsetstrokecolor{textcolor}%
\pgfsetfillcolor{textcolor}%
\pgftext[x=2.050988in,y=0.402469in,,top]{\color{textcolor}\rmfamily\fontsize{10.000000}{12.000000}\selectfont \(\displaystyle 10^{4}\)}%
\end{pgfscope}%
\begin{pgfscope}%
\pgfsetbuttcap%
\pgfsetroundjoin%
\definecolor{currentfill}{rgb}{0.000000,0.000000,0.000000}%
\pgfsetfillcolor{currentfill}%
\pgfsetlinewidth{0.803000pt}%
\definecolor{currentstroke}{rgb}{0.000000,0.000000,0.000000}%
\pgfsetstrokecolor{currentstroke}%
\pgfsetdash{}{0pt}%
\pgfsys@defobject{currentmarker}{\pgfqpoint{0.000000in}{-0.048611in}}{\pgfqpoint{0.000000in}{0.000000in}}{%
\pgfpathmoveto{\pgfqpoint{0.000000in}{0.000000in}}%
\pgfpathlineto{\pgfqpoint{0.000000in}{-0.048611in}}%
\pgfusepath{stroke,fill}%
}%
\begin{pgfscope}%
\pgfsys@transformshift{2.694901in}{0.499691in}%
\pgfsys@useobject{currentmarker}{}%
\end{pgfscope}%
\end{pgfscope}%
\begin{pgfscope}%
\definecolor{textcolor}{rgb}{0.000000,0.000000,0.000000}%
\pgfsetstrokecolor{textcolor}%
\pgfsetfillcolor{textcolor}%
\pgftext[x=2.694901in,y=0.402469in,,top]{\color{textcolor}\rmfamily\fontsize{10.000000}{12.000000}\selectfont \(\displaystyle 10^{5}\)}%
\end{pgfscope}%
\begin{pgfscope}%
\pgfsetbuttcap%
\pgfsetroundjoin%
\definecolor{currentfill}{rgb}{0.000000,0.000000,0.000000}%
\pgfsetfillcolor{currentfill}%
\pgfsetlinewidth{0.602250pt}%
\definecolor{currentstroke}{rgb}{0.000000,0.000000,0.000000}%
\pgfsetstrokecolor{currentstroke}%
\pgfsetdash{}{0pt}%
\pgfsys@defobject{currentmarker}{\pgfqpoint{0.000000in}{-0.027778in}}{\pgfqpoint{0.000000in}{0.000000in}}{%
\pgfpathmoveto{\pgfqpoint{0.000000in}{0.000000in}}%
\pgfpathlineto{\pgfqpoint{0.000000in}{-0.027778in}}%
\pgfusepath{stroke,fill}%
}%
\begin{pgfscope}%
\pgfsys@transformshift{0.663421in}{0.499691in}%
\pgfsys@useobject{currentmarker}{}%
\end{pgfscope}%
\end{pgfscope}%
\begin{pgfscope}%
\pgfsetbuttcap%
\pgfsetroundjoin%
\definecolor{currentfill}{rgb}{0.000000,0.000000,0.000000}%
\pgfsetfillcolor{currentfill}%
\pgfsetlinewidth{0.602250pt}%
\definecolor{currentstroke}{rgb}{0.000000,0.000000,0.000000}%
\pgfsetstrokecolor{currentstroke}%
\pgfsetdash{}{0pt}%
\pgfsys@defobject{currentmarker}{\pgfqpoint{0.000000in}{-0.027778in}}{\pgfqpoint{0.000000in}{0.000000in}}{%
\pgfpathmoveto{\pgfqpoint{0.000000in}{0.000000in}}%
\pgfpathlineto{\pgfqpoint{0.000000in}{-0.027778in}}%
\pgfusepath{stroke,fill}%
}%
\begin{pgfscope}%
\pgfsys@transformshift{0.700762in}{0.499691in}%
\pgfsys@useobject{currentmarker}{}%
\end{pgfscope}%
\end{pgfscope}%
\begin{pgfscope}%
\pgfsetbuttcap%
\pgfsetroundjoin%
\definecolor{currentfill}{rgb}{0.000000,0.000000,0.000000}%
\pgfsetfillcolor{currentfill}%
\pgfsetlinewidth{0.602250pt}%
\definecolor{currentstroke}{rgb}{0.000000,0.000000,0.000000}%
\pgfsetstrokecolor{currentstroke}%
\pgfsetdash{}{0pt}%
\pgfsys@defobject{currentmarker}{\pgfqpoint{0.000000in}{-0.027778in}}{\pgfqpoint{0.000000in}{0.000000in}}{%
\pgfpathmoveto{\pgfqpoint{0.000000in}{0.000000in}}%
\pgfpathlineto{\pgfqpoint{0.000000in}{-0.027778in}}%
\pgfusepath{stroke,fill}%
}%
\begin{pgfscope}%
\pgfsys@transformshift{0.733700in}{0.499691in}%
\pgfsys@useobject{currentmarker}{}%
\end{pgfscope}%
\end{pgfscope}%
\begin{pgfscope}%
\pgfsetbuttcap%
\pgfsetroundjoin%
\definecolor{currentfill}{rgb}{0.000000,0.000000,0.000000}%
\pgfsetfillcolor{currentfill}%
\pgfsetlinewidth{0.602250pt}%
\definecolor{currentstroke}{rgb}{0.000000,0.000000,0.000000}%
\pgfsetstrokecolor{currentstroke}%
\pgfsetdash{}{0pt}%
\pgfsys@defobject{currentmarker}{\pgfqpoint{0.000000in}{-0.027778in}}{\pgfqpoint{0.000000in}{0.000000in}}{%
\pgfpathmoveto{\pgfqpoint{0.000000in}{0.000000in}}%
\pgfpathlineto{\pgfqpoint{0.000000in}{-0.027778in}}%
\pgfusepath{stroke,fill}%
}%
\begin{pgfscope}%
\pgfsys@transformshift{0.957001in}{0.499691in}%
\pgfsys@useobject{currentmarker}{}%
\end{pgfscope}%
\end{pgfscope}%
\begin{pgfscope}%
\pgfsetbuttcap%
\pgfsetroundjoin%
\definecolor{currentfill}{rgb}{0.000000,0.000000,0.000000}%
\pgfsetfillcolor{currentfill}%
\pgfsetlinewidth{0.602250pt}%
\definecolor{currentstroke}{rgb}{0.000000,0.000000,0.000000}%
\pgfsetstrokecolor{currentstroke}%
\pgfsetdash{}{0pt}%
\pgfsys@defobject{currentmarker}{\pgfqpoint{0.000000in}{-0.027778in}}{\pgfqpoint{0.000000in}{0.000000in}}{%
\pgfpathmoveto{\pgfqpoint{0.000000in}{0.000000in}}%
\pgfpathlineto{\pgfqpoint{0.000000in}{-0.027778in}}%
\pgfusepath{stroke,fill}%
}%
\begin{pgfscope}%
\pgfsys@transformshift{1.070388in}{0.499691in}%
\pgfsys@useobject{currentmarker}{}%
\end{pgfscope}%
\end{pgfscope}%
\begin{pgfscope}%
\pgfsetbuttcap%
\pgfsetroundjoin%
\definecolor{currentfill}{rgb}{0.000000,0.000000,0.000000}%
\pgfsetfillcolor{currentfill}%
\pgfsetlinewidth{0.602250pt}%
\definecolor{currentstroke}{rgb}{0.000000,0.000000,0.000000}%
\pgfsetstrokecolor{currentstroke}%
\pgfsetdash{}{0pt}%
\pgfsys@defobject{currentmarker}{\pgfqpoint{0.000000in}{-0.027778in}}{\pgfqpoint{0.000000in}{0.000000in}}{%
\pgfpathmoveto{\pgfqpoint{0.000000in}{0.000000in}}%
\pgfpathlineto{\pgfqpoint{0.000000in}{-0.027778in}}%
\pgfusepath{stroke,fill}%
}%
\begin{pgfscope}%
\pgfsys@transformshift{1.150838in}{0.499691in}%
\pgfsys@useobject{currentmarker}{}%
\end{pgfscope}%
\end{pgfscope}%
\begin{pgfscope}%
\pgfsetbuttcap%
\pgfsetroundjoin%
\definecolor{currentfill}{rgb}{0.000000,0.000000,0.000000}%
\pgfsetfillcolor{currentfill}%
\pgfsetlinewidth{0.602250pt}%
\definecolor{currentstroke}{rgb}{0.000000,0.000000,0.000000}%
\pgfsetstrokecolor{currentstroke}%
\pgfsetdash{}{0pt}%
\pgfsys@defobject{currentmarker}{\pgfqpoint{0.000000in}{-0.027778in}}{\pgfqpoint{0.000000in}{0.000000in}}{%
\pgfpathmoveto{\pgfqpoint{0.000000in}{0.000000in}}%
\pgfpathlineto{\pgfqpoint{0.000000in}{-0.027778in}}%
\pgfusepath{stroke,fill}%
}%
\begin{pgfscope}%
\pgfsys@transformshift{1.213239in}{0.499691in}%
\pgfsys@useobject{currentmarker}{}%
\end{pgfscope}%
\end{pgfscope}%
\begin{pgfscope}%
\pgfsetbuttcap%
\pgfsetroundjoin%
\definecolor{currentfill}{rgb}{0.000000,0.000000,0.000000}%
\pgfsetfillcolor{currentfill}%
\pgfsetlinewidth{0.602250pt}%
\definecolor{currentstroke}{rgb}{0.000000,0.000000,0.000000}%
\pgfsetstrokecolor{currentstroke}%
\pgfsetdash{}{0pt}%
\pgfsys@defobject{currentmarker}{\pgfqpoint{0.000000in}{-0.027778in}}{\pgfqpoint{0.000000in}{0.000000in}}{%
\pgfpathmoveto{\pgfqpoint{0.000000in}{0.000000in}}%
\pgfpathlineto{\pgfqpoint{0.000000in}{-0.027778in}}%
\pgfusepath{stroke,fill}%
}%
\begin{pgfscope}%
\pgfsys@transformshift{1.264225in}{0.499691in}%
\pgfsys@useobject{currentmarker}{}%
\end{pgfscope}%
\end{pgfscope}%
\begin{pgfscope}%
\pgfsetbuttcap%
\pgfsetroundjoin%
\definecolor{currentfill}{rgb}{0.000000,0.000000,0.000000}%
\pgfsetfillcolor{currentfill}%
\pgfsetlinewidth{0.602250pt}%
\definecolor{currentstroke}{rgb}{0.000000,0.000000,0.000000}%
\pgfsetstrokecolor{currentstroke}%
\pgfsetdash{}{0pt}%
\pgfsys@defobject{currentmarker}{\pgfqpoint{0.000000in}{-0.027778in}}{\pgfqpoint{0.000000in}{0.000000in}}{%
\pgfpathmoveto{\pgfqpoint{0.000000in}{0.000000in}}%
\pgfpathlineto{\pgfqpoint{0.000000in}{-0.027778in}}%
\pgfusepath{stroke,fill}%
}%
\begin{pgfscope}%
\pgfsys@transformshift{1.307333in}{0.499691in}%
\pgfsys@useobject{currentmarker}{}%
\end{pgfscope}%
\end{pgfscope}%
\begin{pgfscope}%
\pgfsetbuttcap%
\pgfsetroundjoin%
\definecolor{currentfill}{rgb}{0.000000,0.000000,0.000000}%
\pgfsetfillcolor{currentfill}%
\pgfsetlinewidth{0.602250pt}%
\definecolor{currentstroke}{rgb}{0.000000,0.000000,0.000000}%
\pgfsetstrokecolor{currentstroke}%
\pgfsetdash{}{0pt}%
\pgfsys@defobject{currentmarker}{\pgfqpoint{0.000000in}{-0.027778in}}{\pgfqpoint{0.000000in}{0.000000in}}{%
\pgfpathmoveto{\pgfqpoint{0.000000in}{0.000000in}}%
\pgfpathlineto{\pgfqpoint{0.000000in}{-0.027778in}}%
\pgfusepath{stroke,fill}%
}%
\begin{pgfscope}%
\pgfsys@transformshift{1.344675in}{0.499691in}%
\pgfsys@useobject{currentmarker}{}%
\end{pgfscope}%
\end{pgfscope}%
\begin{pgfscope}%
\pgfsetbuttcap%
\pgfsetroundjoin%
\definecolor{currentfill}{rgb}{0.000000,0.000000,0.000000}%
\pgfsetfillcolor{currentfill}%
\pgfsetlinewidth{0.602250pt}%
\definecolor{currentstroke}{rgb}{0.000000,0.000000,0.000000}%
\pgfsetstrokecolor{currentstroke}%
\pgfsetdash{}{0pt}%
\pgfsys@defobject{currentmarker}{\pgfqpoint{0.000000in}{-0.027778in}}{\pgfqpoint{0.000000in}{0.000000in}}{%
\pgfpathmoveto{\pgfqpoint{0.000000in}{0.000000in}}%
\pgfpathlineto{\pgfqpoint{0.000000in}{-0.027778in}}%
\pgfusepath{stroke,fill}%
}%
\begin{pgfscope}%
\pgfsys@transformshift{1.377612in}{0.499691in}%
\pgfsys@useobject{currentmarker}{}%
\end{pgfscope}%
\end{pgfscope}%
\begin{pgfscope}%
\pgfsetbuttcap%
\pgfsetroundjoin%
\definecolor{currentfill}{rgb}{0.000000,0.000000,0.000000}%
\pgfsetfillcolor{currentfill}%
\pgfsetlinewidth{0.602250pt}%
\definecolor{currentstroke}{rgb}{0.000000,0.000000,0.000000}%
\pgfsetstrokecolor{currentstroke}%
\pgfsetdash{}{0pt}%
\pgfsys@defobject{currentmarker}{\pgfqpoint{0.000000in}{-0.027778in}}{\pgfqpoint{0.000000in}{0.000000in}}{%
\pgfpathmoveto{\pgfqpoint{0.000000in}{0.000000in}}%
\pgfpathlineto{\pgfqpoint{0.000000in}{-0.027778in}}%
\pgfusepath{stroke,fill}%
}%
\begin{pgfscope}%
\pgfsys@transformshift{1.600913in}{0.499691in}%
\pgfsys@useobject{currentmarker}{}%
\end{pgfscope}%
\end{pgfscope}%
\begin{pgfscope}%
\pgfsetbuttcap%
\pgfsetroundjoin%
\definecolor{currentfill}{rgb}{0.000000,0.000000,0.000000}%
\pgfsetfillcolor{currentfill}%
\pgfsetlinewidth{0.602250pt}%
\definecolor{currentstroke}{rgb}{0.000000,0.000000,0.000000}%
\pgfsetstrokecolor{currentstroke}%
\pgfsetdash{}{0pt}%
\pgfsys@defobject{currentmarker}{\pgfqpoint{0.000000in}{-0.027778in}}{\pgfqpoint{0.000000in}{0.000000in}}{%
\pgfpathmoveto{\pgfqpoint{0.000000in}{0.000000in}}%
\pgfpathlineto{\pgfqpoint{0.000000in}{-0.027778in}}%
\pgfusepath{stroke,fill}%
}%
\begin{pgfscope}%
\pgfsys@transformshift{1.714300in}{0.499691in}%
\pgfsys@useobject{currentmarker}{}%
\end{pgfscope}%
\end{pgfscope}%
\begin{pgfscope}%
\pgfsetbuttcap%
\pgfsetroundjoin%
\definecolor{currentfill}{rgb}{0.000000,0.000000,0.000000}%
\pgfsetfillcolor{currentfill}%
\pgfsetlinewidth{0.602250pt}%
\definecolor{currentstroke}{rgb}{0.000000,0.000000,0.000000}%
\pgfsetstrokecolor{currentstroke}%
\pgfsetdash{}{0pt}%
\pgfsys@defobject{currentmarker}{\pgfqpoint{0.000000in}{-0.027778in}}{\pgfqpoint{0.000000in}{0.000000in}}{%
\pgfpathmoveto{\pgfqpoint{0.000000in}{0.000000in}}%
\pgfpathlineto{\pgfqpoint{0.000000in}{-0.027778in}}%
\pgfusepath{stroke,fill}%
}%
\begin{pgfscope}%
\pgfsys@transformshift{1.794750in}{0.499691in}%
\pgfsys@useobject{currentmarker}{}%
\end{pgfscope}%
\end{pgfscope}%
\begin{pgfscope}%
\pgfsetbuttcap%
\pgfsetroundjoin%
\definecolor{currentfill}{rgb}{0.000000,0.000000,0.000000}%
\pgfsetfillcolor{currentfill}%
\pgfsetlinewidth{0.602250pt}%
\definecolor{currentstroke}{rgb}{0.000000,0.000000,0.000000}%
\pgfsetstrokecolor{currentstroke}%
\pgfsetdash{}{0pt}%
\pgfsys@defobject{currentmarker}{\pgfqpoint{0.000000in}{-0.027778in}}{\pgfqpoint{0.000000in}{0.000000in}}{%
\pgfpathmoveto{\pgfqpoint{0.000000in}{0.000000in}}%
\pgfpathlineto{\pgfqpoint{0.000000in}{-0.027778in}}%
\pgfusepath{stroke,fill}%
}%
\begin{pgfscope}%
\pgfsys@transformshift{1.857151in}{0.499691in}%
\pgfsys@useobject{currentmarker}{}%
\end{pgfscope}%
\end{pgfscope}%
\begin{pgfscope}%
\pgfsetbuttcap%
\pgfsetroundjoin%
\definecolor{currentfill}{rgb}{0.000000,0.000000,0.000000}%
\pgfsetfillcolor{currentfill}%
\pgfsetlinewidth{0.602250pt}%
\definecolor{currentstroke}{rgb}{0.000000,0.000000,0.000000}%
\pgfsetstrokecolor{currentstroke}%
\pgfsetdash{}{0pt}%
\pgfsys@defobject{currentmarker}{\pgfqpoint{0.000000in}{-0.027778in}}{\pgfqpoint{0.000000in}{0.000000in}}{%
\pgfpathmoveto{\pgfqpoint{0.000000in}{0.000000in}}%
\pgfpathlineto{\pgfqpoint{0.000000in}{-0.027778in}}%
\pgfusepath{stroke,fill}%
}%
\begin{pgfscope}%
\pgfsys@transformshift{1.908137in}{0.499691in}%
\pgfsys@useobject{currentmarker}{}%
\end{pgfscope}%
\end{pgfscope}%
\begin{pgfscope}%
\pgfsetbuttcap%
\pgfsetroundjoin%
\definecolor{currentfill}{rgb}{0.000000,0.000000,0.000000}%
\pgfsetfillcolor{currentfill}%
\pgfsetlinewidth{0.602250pt}%
\definecolor{currentstroke}{rgb}{0.000000,0.000000,0.000000}%
\pgfsetstrokecolor{currentstroke}%
\pgfsetdash{}{0pt}%
\pgfsys@defobject{currentmarker}{\pgfqpoint{0.000000in}{-0.027778in}}{\pgfqpoint{0.000000in}{0.000000in}}{%
\pgfpathmoveto{\pgfqpoint{0.000000in}{0.000000in}}%
\pgfpathlineto{\pgfqpoint{0.000000in}{-0.027778in}}%
\pgfusepath{stroke,fill}%
}%
\begin{pgfscope}%
\pgfsys@transformshift{1.951245in}{0.499691in}%
\pgfsys@useobject{currentmarker}{}%
\end{pgfscope}%
\end{pgfscope}%
\begin{pgfscope}%
\pgfsetbuttcap%
\pgfsetroundjoin%
\definecolor{currentfill}{rgb}{0.000000,0.000000,0.000000}%
\pgfsetfillcolor{currentfill}%
\pgfsetlinewidth{0.602250pt}%
\definecolor{currentstroke}{rgb}{0.000000,0.000000,0.000000}%
\pgfsetstrokecolor{currentstroke}%
\pgfsetdash{}{0pt}%
\pgfsys@defobject{currentmarker}{\pgfqpoint{0.000000in}{-0.027778in}}{\pgfqpoint{0.000000in}{0.000000in}}{%
\pgfpathmoveto{\pgfqpoint{0.000000in}{0.000000in}}%
\pgfpathlineto{\pgfqpoint{0.000000in}{-0.027778in}}%
\pgfusepath{stroke,fill}%
}%
\begin{pgfscope}%
\pgfsys@transformshift{1.988587in}{0.499691in}%
\pgfsys@useobject{currentmarker}{}%
\end{pgfscope}%
\end{pgfscope}%
\begin{pgfscope}%
\pgfsetbuttcap%
\pgfsetroundjoin%
\definecolor{currentfill}{rgb}{0.000000,0.000000,0.000000}%
\pgfsetfillcolor{currentfill}%
\pgfsetlinewidth{0.602250pt}%
\definecolor{currentstroke}{rgb}{0.000000,0.000000,0.000000}%
\pgfsetstrokecolor{currentstroke}%
\pgfsetdash{}{0pt}%
\pgfsys@defobject{currentmarker}{\pgfqpoint{0.000000in}{-0.027778in}}{\pgfqpoint{0.000000in}{0.000000in}}{%
\pgfpathmoveto{\pgfqpoint{0.000000in}{0.000000in}}%
\pgfpathlineto{\pgfqpoint{0.000000in}{-0.027778in}}%
\pgfusepath{stroke,fill}%
}%
\begin{pgfscope}%
\pgfsys@transformshift{2.021525in}{0.499691in}%
\pgfsys@useobject{currentmarker}{}%
\end{pgfscope}%
\end{pgfscope}%
\begin{pgfscope}%
\pgfsetbuttcap%
\pgfsetroundjoin%
\definecolor{currentfill}{rgb}{0.000000,0.000000,0.000000}%
\pgfsetfillcolor{currentfill}%
\pgfsetlinewidth{0.602250pt}%
\definecolor{currentstroke}{rgb}{0.000000,0.000000,0.000000}%
\pgfsetstrokecolor{currentstroke}%
\pgfsetdash{}{0pt}%
\pgfsys@defobject{currentmarker}{\pgfqpoint{0.000000in}{-0.027778in}}{\pgfqpoint{0.000000in}{0.000000in}}{%
\pgfpathmoveto{\pgfqpoint{0.000000in}{0.000000in}}%
\pgfpathlineto{\pgfqpoint{0.000000in}{-0.027778in}}%
\pgfusepath{stroke,fill}%
}%
\begin{pgfscope}%
\pgfsys@transformshift{2.244825in}{0.499691in}%
\pgfsys@useobject{currentmarker}{}%
\end{pgfscope}%
\end{pgfscope}%
\begin{pgfscope}%
\pgfsetbuttcap%
\pgfsetroundjoin%
\definecolor{currentfill}{rgb}{0.000000,0.000000,0.000000}%
\pgfsetfillcolor{currentfill}%
\pgfsetlinewidth{0.602250pt}%
\definecolor{currentstroke}{rgb}{0.000000,0.000000,0.000000}%
\pgfsetstrokecolor{currentstroke}%
\pgfsetdash{}{0pt}%
\pgfsys@defobject{currentmarker}{\pgfqpoint{0.000000in}{-0.027778in}}{\pgfqpoint{0.000000in}{0.000000in}}{%
\pgfpathmoveto{\pgfqpoint{0.000000in}{0.000000in}}%
\pgfpathlineto{\pgfqpoint{0.000000in}{-0.027778in}}%
\pgfusepath{stroke,fill}%
}%
\begin{pgfscope}%
\pgfsys@transformshift{2.358213in}{0.499691in}%
\pgfsys@useobject{currentmarker}{}%
\end{pgfscope}%
\end{pgfscope}%
\begin{pgfscope}%
\pgfsetbuttcap%
\pgfsetroundjoin%
\definecolor{currentfill}{rgb}{0.000000,0.000000,0.000000}%
\pgfsetfillcolor{currentfill}%
\pgfsetlinewidth{0.602250pt}%
\definecolor{currentstroke}{rgb}{0.000000,0.000000,0.000000}%
\pgfsetstrokecolor{currentstroke}%
\pgfsetdash{}{0pt}%
\pgfsys@defobject{currentmarker}{\pgfqpoint{0.000000in}{-0.027778in}}{\pgfqpoint{0.000000in}{0.000000in}}{%
\pgfpathmoveto{\pgfqpoint{0.000000in}{0.000000in}}%
\pgfpathlineto{\pgfqpoint{0.000000in}{-0.027778in}}%
\pgfusepath{stroke,fill}%
}%
\begin{pgfscope}%
\pgfsys@transformshift{2.438662in}{0.499691in}%
\pgfsys@useobject{currentmarker}{}%
\end{pgfscope}%
\end{pgfscope}%
\begin{pgfscope}%
\pgfsetbuttcap%
\pgfsetroundjoin%
\definecolor{currentfill}{rgb}{0.000000,0.000000,0.000000}%
\pgfsetfillcolor{currentfill}%
\pgfsetlinewidth{0.602250pt}%
\definecolor{currentstroke}{rgb}{0.000000,0.000000,0.000000}%
\pgfsetstrokecolor{currentstroke}%
\pgfsetdash{}{0pt}%
\pgfsys@defobject{currentmarker}{\pgfqpoint{0.000000in}{-0.027778in}}{\pgfqpoint{0.000000in}{0.000000in}}{%
\pgfpathmoveto{\pgfqpoint{0.000000in}{0.000000in}}%
\pgfpathlineto{\pgfqpoint{0.000000in}{-0.027778in}}%
\pgfusepath{stroke,fill}%
}%
\begin{pgfscope}%
\pgfsys@transformshift{2.501064in}{0.499691in}%
\pgfsys@useobject{currentmarker}{}%
\end{pgfscope}%
\end{pgfscope}%
\begin{pgfscope}%
\pgfsetbuttcap%
\pgfsetroundjoin%
\definecolor{currentfill}{rgb}{0.000000,0.000000,0.000000}%
\pgfsetfillcolor{currentfill}%
\pgfsetlinewidth{0.602250pt}%
\definecolor{currentstroke}{rgb}{0.000000,0.000000,0.000000}%
\pgfsetstrokecolor{currentstroke}%
\pgfsetdash{}{0pt}%
\pgfsys@defobject{currentmarker}{\pgfqpoint{0.000000in}{-0.027778in}}{\pgfqpoint{0.000000in}{0.000000in}}{%
\pgfpathmoveto{\pgfqpoint{0.000000in}{0.000000in}}%
\pgfpathlineto{\pgfqpoint{0.000000in}{-0.027778in}}%
\pgfusepath{stroke,fill}%
}%
\begin{pgfscope}%
\pgfsys@transformshift{2.552049in}{0.499691in}%
\pgfsys@useobject{currentmarker}{}%
\end{pgfscope}%
\end{pgfscope}%
\begin{pgfscope}%
\pgfsetbuttcap%
\pgfsetroundjoin%
\definecolor{currentfill}{rgb}{0.000000,0.000000,0.000000}%
\pgfsetfillcolor{currentfill}%
\pgfsetlinewidth{0.602250pt}%
\definecolor{currentstroke}{rgb}{0.000000,0.000000,0.000000}%
\pgfsetstrokecolor{currentstroke}%
\pgfsetdash{}{0pt}%
\pgfsys@defobject{currentmarker}{\pgfqpoint{0.000000in}{-0.027778in}}{\pgfqpoint{0.000000in}{0.000000in}}{%
\pgfpathmoveto{\pgfqpoint{0.000000in}{0.000000in}}%
\pgfpathlineto{\pgfqpoint{0.000000in}{-0.027778in}}%
\pgfusepath{stroke,fill}%
}%
\begin{pgfscope}%
\pgfsys@transformshift{2.595157in}{0.499691in}%
\pgfsys@useobject{currentmarker}{}%
\end{pgfscope}%
\end{pgfscope}%
\begin{pgfscope}%
\pgfsetbuttcap%
\pgfsetroundjoin%
\definecolor{currentfill}{rgb}{0.000000,0.000000,0.000000}%
\pgfsetfillcolor{currentfill}%
\pgfsetlinewidth{0.602250pt}%
\definecolor{currentstroke}{rgb}{0.000000,0.000000,0.000000}%
\pgfsetstrokecolor{currentstroke}%
\pgfsetdash{}{0pt}%
\pgfsys@defobject{currentmarker}{\pgfqpoint{0.000000in}{-0.027778in}}{\pgfqpoint{0.000000in}{0.000000in}}{%
\pgfpathmoveto{\pgfqpoint{0.000000in}{0.000000in}}%
\pgfpathlineto{\pgfqpoint{0.000000in}{-0.027778in}}%
\pgfusepath{stroke,fill}%
}%
\begin{pgfscope}%
\pgfsys@transformshift{2.632499in}{0.499691in}%
\pgfsys@useobject{currentmarker}{}%
\end{pgfscope}%
\end{pgfscope}%
\begin{pgfscope}%
\pgfsetbuttcap%
\pgfsetroundjoin%
\definecolor{currentfill}{rgb}{0.000000,0.000000,0.000000}%
\pgfsetfillcolor{currentfill}%
\pgfsetlinewidth{0.602250pt}%
\definecolor{currentstroke}{rgb}{0.000000,0.000000,0.000000}%
\pgfsetstrokecolor{currentstroke}%
\pgfsetdash{}{0pt}%
\pgfsys@defobject{currentmarker}{\pgfqpoint{0.000000in}{-0.027778in}}{\pgfqpoint{0.000000in}{0.000000in}}{%
\pgfpathmoveto{\pgfqpoint{0.000000in}{0.000000in}}%
\pgfpathlineto{\pgfqpoint{0.000000in}{-0.027778in}}%
\pgfusepath{stroke,fill}%
}%
\begin{pgfscope}%
\pgfsys@transformshift{2.665437in}{0.499691in}%
\pgfsys@useobject{currentmarker}{}%
\end{pgfscope}%
\end{pgfscope}%
\begin{pgfscope}%
\definecolor{textcolor}{rgb}{0.000000,0.000000,0.000000}%
\pgfsetstrokecolor{textcolor}%
\pgfsetfillcolor{textcolor}%
\pgftext[x=1.684831in,y=0.223457in,,top]{\color{textcolor}\rmfamily\fontsize{10.000000}{12.000000}\selectfont no. of shapes}%
\end{pgfscope}%
\begin{pgfscope}%
\pgfsetbuttcap%
\pgfsetroundjoin%
\definecolor{currentfill}{rgb}{0.000000,0.000000,0.000000}%
\pgfsetfillcolor{currentfill}%
\pgfsetlinewidth{0.803000pt}%
\definecolor{currentstroke}{rgb}{0.000000,0.000000,0.000000}%
\pgfsetstrokecolor{currentstroke}%
\pgfsetdash{}{0pt}%
\pgfsys@defobject{currentmarker}{\pgfqpoint{-0.048611in}{0.000000in}}{\pgfqpoint{0.000000in}{0.000000in}}{%
\pgfpathmoveto{\pgfqpoint{0.000000in}{0.000000in}}%
\pgfpathlineto{\pgfqpoint{-0.048611in}{0.000000in}}%
\pgfusepath{stroke,fill}%
}%
\begin{pgfscope}%
\pgfsys@transformshift{0.638581in}{0.602646in}%
\pgfsys@useobject{currentmarker}{}%
\end{pgfscope}%
\end{pgfscope}%
\begin{pgfscope}%
\definecolor{textcolor}{rgb}{0.000000,0.000000,0.000000}%
\pgfsetstrokecolor{textcolor}%
\pgfsetfillcolor{textcolor}%
\pgftext[x=0.294444in,y=0.554420in,left,base]{\color{textcolor}\rmfamily\fontsize{10.000000}{12.000000}\selectfont \(\displaystyle 0.02\)}%
\end{pgfscope}%
\begin{pgfscope}%
\pgfsetbuttcap%
\pgfsetroundjoin%
\definecolor{currentfill}{rgb}{0.000000,0.000000,0.000000}%
\pgfsetfillcolor{currentfill}%
\pgfsetlinewidth{0.803000pt}%
\definecolor{currentstroke}{rgb}{0.000000,0.000000,0.000000}%
\pgfsetstrokecolor{currentstroke}%
\pgfsetdash{}{0pt}%
\pgfsys@defobject{currentmarker}{\pgfqpoint{-0.048611in}{0.000000in}}{\pgfqpoint{0.000000in}{0.000000in}}{%
\pgfpathmoveto{\pgfqpoint{0.000000in}{0.000000in}}%
\pgfpathlineto{\pgfqpoint{-0.048611in}{0.000000in}}%
\pgfusepath{stroke,fill}%
}%
\begin{pgfscope}%
\pgfsys@transformshift{0.638581in}{0.994853in}%
\pgfsys@useobject{currentmarker}{}%
\end{pgfscope}%
\end{pgfscope}%
\begin{pgfscope}%
\definecolor{textcolor}{rgb}{0.000000,0.000000,0.000000}%
\pgfsetstrokecolor{textcolor}%
\pgfsetfillcolor{textcolor}%
\pgftext[x=0.294444in,y=0.946628in,left,base]{\color{textcolor}\rmfamily\fontsize{10.000000}{12.000000}\selectfont \(\displaystyle 0.04\)}%
\end{pgfscope}%
\begin{pgfscope}%
\pgfsetbuttcap%
\pgfsetroundjoin%
\definecolor{currentfill}{rgb}{0.000000,0.000000,0.000000}%
\pgfsetfillcolor{currentfill}%
\pgfsetlinewidth{0.803000pt}%
\definecolor{currentstroke}{rgb}{0.000000,0.000000,0.000000}%
\pgfsetstrokecolor{currentstroke}%
\pgfsetdash{}{0pt}%
\pgfsys@defobject{currentmarker}{\pgfqpoint{-0.048611in}{0.000000in}}{\pgfqpoint{0.000000in}{0.000000in}}{%
\pgfpathmoveto{\pgfqpoint{0.000000in}{0.000000in}}%
\pgfpathlineto{\pgfqpoint{-0.048611in}{0.000000in}}%
\pgfusepath{stroke,fill}%
}%
\begin{pgfscope}%
\pgfsys@transformshift{0.638581in}{1.387061in}%
\pgfsys@useobject{currentmarker}{}%
\end{pgfscope}%
\end{pgfscope}%
\begin{pgfscope}%
\definecolor{textcolor}{rgb}{0.000000,0.000000,0.000000}%
\pgfsetstrokecolor{textcolor}%
\pgfsetfillcolor{textcolor}%
\pgftext[x=0.294444in,y=1.338836in,left,base]{\color{textcolor}\rmfamily\fontsize{10.000000}{12.000000}\selectfont \(\displaystyle 0.06\)}%
\end{pgfscope}%
\begin{pgfscope}%
\pgfsetbuttcap%
\pgfsetroundjoin%
\definecolor{currentfill}{rgb}{0.000000,0.000000,0.000000}%
\pgfsetfillcolor{currentfill}%
\pgfsetlinewidth{0.803000pt}%
\definecolor{currentstroke}{rgb}{0.000000,0.000000,0.000000}%
\pgfsetstrokecolor{currentstroke}%
\pgfsetdash{}{0pt}%
\pgfsys@defobject{currentmarker}{\pgfqpoint{-0.048611in}{0.000000in}}{\pgfqpoint{0.000000in}{0.000000in}}{%
\pgfpathmoveto{\pgfqpoint{0.000000in}{0.000000in}}%
\pgfpathlineto{\pgfqpoint{-0.048611in}{0.000000in}}%
\pgfusepath{stroke,fill}%
}%
\begin{pgfscope}%
\pgfsys@transformshift{0.638581in}{1.779269in}%
\pgfsys@useobject{currentmarker}{}%
\end{pgfscope}%
\end{pgfscope}%
\begin{pgfscope}%
\definecolor{textcolor}{rgb}{0.000000,0.000000,0.000000}%
\pgfsetstrokecolor{textcolor}%
\pgfsetfillcolor{textcolor}%
\pgftext[x=0.294444in,y=1.731044in,left,base]{\color{textcolor}\rmfamily\fontsize{10.000000}{12.000000}\selectfont \(\displaystyle 0.08\)}%
\end{pgfscope}%
\begin{pgfscope}%
\pgfsetbuttcap%
\pgfsetroundjoin%
\definecolor{currentfill}{rgb}{0.000000,0.000000,0.000000}%
\pgfsetfillcolor{currentfill}%
\pgfsetlinewidth{0.803000pt}%
\definecolor{currentstroke}{rgb}{0.000000,0.000000,0.000000}%
\pgfsetstrokecolor{currentstroke}%
\pgfsetdash{}{0pt}%
\pgfsys@defobject{currentmarker}{\pgfqpoint{-0.048611in}{0.000000in}}{\pgfqpoint{0.000000in}{0.000000in}}{%
\pgfpathmoveto{\pgfqpoint{0.000000in}{0.000000in}}%
\pgfpathlineto{\pgfqpoint{-0.048611in}{0.000000in}}%
\pgfusepath{stroke,fill}%
}%
\begin{pgfscope}%
\pgfsys@transformshift{0.638581in}{2.171477in}%
\pgfsys@useobject{currentmarker}{}%
\end{pgfscope}%
\end{pgfscope}%
\begin{pgfscope}%
\definecolor{textcolor}{rgb}{0.000000,0.000000,0.000000}%
\pgfsetstrokecolor{textcolor}%
\pgfsetfillcolor{textcolor}%
\pgftext[x=0.294444in,y=2.123252in,left,base]{\color{textcolor}\rmfamily\fontsize{10.000000}{12.000000}\selectfont \(\displaystyle 0.10\)}%
\end{pgfscope}%
\begin{pgfscope}%
\pgfsetbuttcap%
\pgfsetroundjoin%
\definecolor{currentfill}{rgb}{0.000000,0.000000,0.000000}%
\pgfsetfillcolor{currentfill}%
\pgfsetlinewidth{0.803000pt}%
\definecolor{currentstroke}{rgb}{0.000000,0.000000,0.000000}%
\pgfsetstrokecolor{currentstroke}%
\pgfsetdash{}{0pt}%
\pgfsys@defobject{currentmarker}{\pgfqpoint{-0.048611in}{0.000000in}}{\pgfqpoint{0.000000in}{0.000000in}}{%
\pgfpathmoveto{\pgfqpoint{0.000000in}{0.000000in}}%
\pgfpathlineto{\pgfqpoint{-0.048611in}{0.000000in}}%
\pgfusepath{stroke,fill}%
}%
\begin{pgfscope}%
\pgfsys@transformshift{0.638581in}{2.563685in}%
\pgfsys@useobject{currentmarker}{}%
\end{pgfscope}%
\end{pgfscope}%
\begin{pgfscope}%
\definecolor{textcolor}{rgb}{0.000000,0.000000,0.000000}%
\pgfsetstrokecolor{textcolor}%
\pgfsetfillcolor{textcolor}%
\pgftext[x=0.294444in,y=2.515459in,left,base]{\color{textcolor}\rmfamily\fontsize{10.000000}{12.000000}\selectfont \(\displaystyle 0.12\)}%
\end{pgfscope}%
\begin{pgfscope}%
\definecolor{textcolor}{rgb}{0.000000,0.000000,0.000000}%
\pgfsetstrokecolor{textcolor}%
\pgfsetfillcolor{textcolor}%
\pgftext[x=0.238889in,y=1.632191in,,bottom,rotate=90.000000]{\color{textcolor}\rmfamily\fontsize{10.000000}{12.000000}\selectfont MSE/\(\displaystyle \sigma\)}%
\end{pgfscope}%
\begin{pgfscope}%
\pgfpathrectangle{\pgfqpoint{0.638581in}{0.499691in}}{\pgfqpoint{2.092500in}{2.265000in}}%
\pgfusepath{clip}%
\pgfsetbuttcap%
\pgfsetroundjoin%
\pgfsetlinewidth{1.505625pt}%
\definecolor{currentstroke}{rgb}{0.000000,0.000000,0.000000}%
\pgfsetstrokecolor{currentstroke}%
\pgfsetdash{{5.550000pt}{2.400000pt}}{0.000000pt}%
\pgfpathmoveto{\pgfqpoint{2.635967in}{0.602646in}}%
\pgfpathlineto{\pgfqpoint{2.471594in}{0.700698in}}%
\pgfpathlineto{\pgfqpoint{2.215356in}{0.700698in}}%
\pgfpathlineto{\pgfqpoint{2.021519in}{0.700698in}}%
\pgfpathlineto{\pgfqpoint{1.827682in}{0.739918in}}%
\pgfpathlineto{\pgfqpoint{1.571444in}{0.837970in}}%
\pgfpathlineto{\pgfqpoint{1.377607in}{0.857581in}}%
\pgfpathlineto{\pgfqpoint{0.927537in}{1.190957in}}%
\pgfpathlineto{\pgfqpoint{0.733695in}{2.661737in}}%
\pgfusepath{stroke}%
\end{pgfscope}%
\begin{pgfscope}%
\pgfpathrectangle{\pgfqpoint{0.638581in}{0.499691in}}{\pgfqpoint{2.092500in}{2.265000in}}%
\pgfusepath{clip}%
\pgfsetbuttcap%
\pgfsetroundjoin%
\definecolor{currentfill}{rgb}{0.000000,0.000000,0.000000}%
\pgfsetfillcolor{currentfill}%
\pgfsetlinewidth{1.003750pt}%
\definecolor{currentstroke}{rgb}{0.000000,0.000000,0.000000}%
\pgfsetstrokecolor{currentstroke}%
\pgfsetdash{}{0pt}%
\pgfsys@defobject{currentmarker}{\pgfqpoint{-0.041667in}{-0.041667in}}{\pgfqpoint{0.041667in}{0.041667in}}{%
\pgfpathmoveto{\pgfqpoint{0.000000in}{-0.041667in}}%
\pgfpathcurveto{\pgfqpoint{0.011050in}{-0.041667in}}{\pgfqpoint{0.021649in}{-0.037276in}}{\pgfqpoint{0.029463in}{-0.029463in}}%
\pgfpathcurveto{\pgfqpoint{0.037276in}{-0.021649in}}{\pgfqpoint{0.041667in}{-0.011050in}}{\pgfqpoint{0.041667in}{0.000000in}}%
\pgfpathcurveto{\pgfqpoint{0.041667in}{0.011050in}}{\pgfqpoint{0.037276in}{0.021649in}}{\pgfqpoint{0.029463in}{0.029463in}}%
\pgfpathcurveto{\pgfqpoint{0.021649in}{0.037276in}}{\pgfqpoint{0.011050in}{0.041667in}}{\pgfqpoint{0.000000in}{0.041667in}}%
\pgfpathcurveto{\pgfqpoint{-0.011050in}{0.041667in}}{\pgfqpoint{-0.021649in}{0.037276in}}{\pgfqpoint{-0.029463in}{0.029463in}}%
\pgfpathcurveto{\pgfqpoint{-0.037276in}{0.021649in}}{\pgfqpoint{-0.041667in}{0.011050in}}{\pgfqpoint{-0.041667in}{0.000000in}}%
\pgfpathcurveto{\pgfqpoint{-0.041667in}{-0.011050in}}{\pgfqpoint{-0.037276in}{-0.021649in}}{\pgfqpoint{-0.029463in}{-0.029463in}}%
\pgfpathcurveto{\pgfqpoint{-0.021649in}{-0.037276in}}{\pgfqpoint{-0.011050in}{-0.041667in}}{\pgfqpoint{0.000000in}{-0.041667in}}%
\pgfpathclose%
\pgfusepath{stroke,fill}%
}%
\begin{pgfscope}%
\pgfsys@transformshift{2.635967in}{0.602646in}%
\pgfsys@useobject{currentmarker}{}%
\end{pgfscope}%
\begin{pgfscope}%
\pgfsys@transformshift{2.471594in}{0.700698in}%
\pgfsys@useobject{currentmarker}{}%
\end{pgfscope}%
\begin{pgfscope}%
\pgfsys@transformshift{2.215356in}{0.700698in}%
\pgfsys@useobject{currentmarker}{}%
\end{pgfscope}%
\begin{pgfscope}%
\pgfsys@transformshift{2.021519in}{0.700698in}%
\pgfsys@useobject{currentmarker}{}%
\end{pgfscope}%
\begin{pgfscope}%
\pgfsys@transformshift{1.827682in}{0.739918in}%
\pgfsys@useobject{currentmarker}{}%
\end{pgfscope}%
\begin{pgfscope}%
\pgfsys@transformshift{1.571444in}{0.837970in}%
\pgfsys@useobject{currentmarker}{}%
\end{pgfscope}%
\begin{pgfscope}%
\pgfsys@transformshift{1.377607in}{0.857581in}%
\pgfsys@useobject{currentmarker}{}%
\end{pgfscope}%
\begin{pgfscope}%
\pgfsys@transformshift{0.927537in}{1.190957in}%
\pgfsys@useobject{currentmarker}{}%
\end{pgfscope}%
\begin{pgfscope}%
\pgfsys@transformshift{0.733695in}{2.661737in}%
\pgfsys@useobject{currentmarker}{}%
\end{pgfscope}%
\end{pgfscope}%
\begin{pgfscope}%
\pgfsetrectcap%
\pgfsetmiterjoin%
\pgfsetlinewidth{0.803000pt}%
\definecolor{currentstroke}{rgb}{0.000000,0.000000,0.000000}%
\pgfsetstrokecolor{currentstroke}%
\pgfsetdash{}{0pt}%
\pgfpathmoveto{\pgfqpoint{0.638581in}{0.499691in}}%
\pgfpathlineto{\pgfqpoint{0.638581in}{2.764691in}}%
\pgfusepath{stroke}%
\end{pgfscope}%
\begin{pgfscope}%
\pgfsetrectcap%
\pgfsetmiterjoin%
\pgfsetlinewidth{0.803000pt}%
\definecolor{currentstroke}{rgb}{0.000000,0.000000,0.000000}%
\pgfsetstrokecolor{currentstroke}%
\pgfsetdash{}{0pt}%
\pgfpathmoveto{\pgfqpoint{2.731081in}{0.499691in}}%
\pgfpathlineto{\pgfqpoint{2.731081in}{2.764691in}}%
\pgfusepath{stroke}%
\end{pgfscope}%
\begin{pgfscope}%
\pgfsetrectcap%
\pgfsetmiterjoin%
\pgfsetlinewidth{0.803000pt}%
\definecolor{currentstroke}{rgb}{0.000000,0.000000,0.000000}%
\pgfsetstrokecolor{currentstroke}%
\pgfsetdash{}{0pt}%
\pgfpathmoveto{\pgfqpoint{0.638581in}{0.499691in}}%
\pgfpathlineto{\pgfqpoint{2.731081in}{0.499691in}}%
\pgfusepath{stroke}%
\end{pgfscope}%
\begin{pgfscope}%
\pgfsetrectcap%
\pgfsetmiterjoin%
\pgfsetlinewidth{0.803000pt}%
\definecolor{currentstroke}{rgb}{0.000000,0.000000,0.000000}%
\pgfsetstrokecolor{currentstroke}%
\pgfsetdash{}{0pt}%
\pgfpathmoveto{\pgfqpoint{0.638581in}{2.764691in}}%
\pgfpathlineto{\pgfqpoint{2.731081in}{2.764691in}}%
\pgfusepath{stroke}%
\end{pgfscope}%
\end{pgfpicture}%
\makeatother%
\endgroup%